\documentclass[12pt]{article}
\usepackage{amsmath}
\usepackage{times}
\usepackage{graphicx}
\usepackage{color}
\usepackage{multirow}
\usepackage[authoryear]{natbib}
\usepackage{rotating}
\usepackage{bbm}
\usepackage{latexsym}

\newcommand{\captionfonts}{\normalsize}

\makeatletter  
\long\def\@makecaption#1#2{%
  \vskip\abovecaptionskip
  \sbox\@tempboxa{{\captionfonts #1: #2}}%
  \ifdim \wd\@tempboxa >\hsize
    {\captionfonts #1: #2\par}
  \else
    \hbox to\hsize{\hfil\box\@tempboxa\hfil}%
  \fi
  \vskip\belowcaptionskip}
\makeatother   

\usepackage{hyperref}       
\usepackage{url}            
\usepackage{booktabs}       
\usepackage{amsfonts}       
\usepackage{nicefrac}       
\usepackage{microtype}      
\usepackage{xcolor}         
\usepackage{graphicx}
\usepackage{amsmath}
\usepackage{gensymb}
\usepackage{amssymb}
\usepackage{subcaption}
\usepackage[ruled,vlined]{algorithm2e}
\usepackage{wrapfig}

\newcommand{\mtx}[1]{\ensuremath{\mathbf{#1}}}
\newcommand{\vtr}[1]{\ensuremath{\mathbf{#1}}}

\date{}

\begin{document}

\hspace{13.9cm}1

\ \vspace{20mm}\\

{\LARGE Learning Internal Representations of 3D Transformations from 2D Projected Inputs}

\ \\
{\bf \large Marissa Connor$^{\displaystyle 1}$ Bruno Olshausen$^{\displaystyle 2}$  Christopher Rozell$^{\displaystyle 1}$}\\
{$^{\displaystyle 1}$School of Electrical and Computer Engineering,
Georgia Institute of Technology}\\
{$^{\displaystyle 2}$Helen Wills Neuroscience Institute and
School of Optometry, University of California, Berkeley}\\
\\


\begin{center} {\bf Abstract} \end{center}

When interacting in a three dimensional world, humans must estimate 3D structure from visual inputs projected down to two dimensional retinal images. It has been shown that humans use the persistence of object shape over motion-induced transformations as a cue to resolve depth ambiguity when solving this underconstrained problem. With the aim of understanding how biological vision systems may internally represent 3D transformations, we propose a computational model, based on a generative manifold model, which can be used to infer 3D structure from the motion of 2D points. Our model can also learn representations of the transformations with minimal supervision, providing a proof of concept for how humans may develop internal representations on a developmental or evolutionary time scale. Focused on rotational motion, we show how our model infers depth from moving 2D projected points, learns 3D rotational transformations from 2D training stimuli, and compares to human performance on psychophysical structure-from-motion experiments. 

\section{Introduction}

For effective interactions with 3D objects and environments, humans must estimate 3D structure from visual inputs projected down to two dimensions in a retinal image. This problem of recovering 3D structure from 2D projections is underconstrained -- there are an infinite number of possible depths for each 2D point. To resolve this challenge, humans rely on a variety of cues to infer depth including motion parallax, binocular disparity, texture, occlusions, shadows, size, blur, and shading~\citep{reichelt2010depth}. Specifically, the persistence of object shape over motion-induced transformations provides a powerful cue (even in the absence of other cues~\citep{petersik1979three,braunstein1987minimum,todd1988apparent,dosher1989ratings,sperling1989kinetic,braunstein2014depth}) that can be used to resolve the depth ambiguity  for points on an object's surface and improve accuracy of depth perception.

Mathematically there are precise definitions of motion-induced geometric transformations such as rotations and translations, and these can be employed to successfully compute point depths from multiple viewpoints~\citep{longuet1981computer,fischler1981random,tomasi1992shape,nister2005preemptive,pollefeys2008detailed,snavely2006photo} or frames~\citep{godard2017unsupervised,garg2016unsupervised,xie2016deep3d,zhou2017unsupervised}. While all of these methods use analytically constructed models to achieve the objective of estimating depth from multiple viewpoints, we aim to understand how biological vision systems may internally represent 3D transformations, how they could learn or adapt to the statistics of these 3D transformations with minimal supervision, and how this knowledge could be used to aid in discerning structure from motion. The results of mental transformation experiments~\citep{shepard1986mental,lamm2007functional}, as well as qualitative descriptions from subjects performing mental transformation tasks~\citep{zacks2005transformations}, suggest that humans internally imagine 3D spatial transformations when performing tasks such as identifying rotated reference objects~\citep{shepard1971mental,cooper1973chronometric,just1985cognitive}. We propose a computational model to explain the mechanism for internally representing these transformations and for learning them with minimal supervision~\citep{perry2010continuous} and show that this model can be used to infer 3D structure from the motion 2D points. 

Motivated by the manifold hypothesis which states that natural variations in high-dimensional data lie on or near a low-dimensional, nonlinear manifold~\citep{fefferman2016testing}, we introduce generative manifold models as a possible mechanism for learning and representing internal models of natural motion-induced transformations. These models represent manifolds through continuous, nonlinear transformation operators that traverse the geometric structure of the manifold~\citep{culpepper2009learning,sohl2010unsupervised,connor2020representing,connor2021variational,connor2021learning}. The transformation operators can be used to infer relationships between different object views and to interpolate or extrapolate views of transformed objects. Neuroscience research has suggested that the brain explicitly exploits the manifold structure of object variations by using hierarchical processing stages to flatten the manifolds produced by different objects undergoing the same physical transformations (e.g., changes in pose and position)~\citep{dicarlo2007untangling,dicarlo2012does}, but to our knowledge no detailed model has been proposed for how a biological system could learn or represent the manifolds of such natural variations from data. 

Specifically in this work, we develop a proof of concept for the viability of learning 3D transformation representations from 2D projected inputs using a generative manifold model. Focusing on the rotational motion that is used in many structure from motion tasks~\citep{petersik1979three,dosher1989ratings,braunstein2014depth}, we develop a manifold-based method for inferring depth from moving 2D projected points and learning 3D rotational transformation models from 2D training stimuli. Finally, we apply the learned transformation model to structure from motion tasks and compare to human performance on psychophysical experiments.

\section{Background}

In this work, we focus on the development of a model that can learn transformations which may be used to model internal mental rotation. While there have been computational models introduced for mental rotation~\citep{funt1983parallel,fukumi1997rotation,inui2011temporo,seepanomwan2015generalisation}, they have both assumed prior knowledge of the rotational transformations and been focused on modeling specific brain areas that are involved in this process. In contrast, we  focus on the representation of the transformation model itself, including the learning and inference process within such a model.

We use our model of 3D transformations to infer point depths from 2D projections of moving points. The ability for humans to perceive depth from moving points and objects, known as the kinetic depth effect~\citep{wallach1953kinetic}, has been extensively studied in both psychology and computer vision. The kinetic depth effect has been investigated through a wide array of psychophysical experiments suggesting that humans can generate stable precepts of 3D structures under a wide variety of conditions~\citep{petersik1979three,braunstein1987minimum,todd1988apparent,dosher1989ratings,sperling1989kinetic,braunstein2014depth}.

Computational models have been developed to estimate 3D point-cloud structure from multiple views of an object or scene using multiview geometry~\citep{longuet1981computer,tsai1984uniqueness,hartley1997defense,hartley1997triangulation,Hartley2004}, factorization methods~\citep{tomasi1992shape,kanade1998factorization}, and neural network-based models~\citep{eigen2014depth,ladicky2014pulling,liu2015learning,godard2017unsupervised,garg2016unsupervised,xie2016deep3d,godard2017unsupervised,garg2016unsupervised,xie2016deep3d}. All of these methods, while very successful at estimating camera motion and depth, have requirements that make them poor representations of the neural mechanisms for learning types of motion, inferring motion in scenes, and estimating point depths. Several methods assume prior knowledge of the types of transformations present in temporal visual inputs (i.e., rotation and translation) and rely on a mathematical specification of how to apply rotational and translational motion through matrix multiplication. Other methods require ground truth depth labels in order to train a system to estimate depth and motion. We cannot assume that vision systems know ground truth point depths or how to apply natural transformations during the development of mechanisms for estimating motion and depth of scenes. In this work, we learn a representation of 3D transformations from observed point motion itself, without a prior assumption how these transformations affect points in a scene. This could resemble how internal mental transformation models are developed. Importantly, this model is learned using only 2D moving points without requiring ground truth knowledge of point depth.

\section{Model Description}\label{sec:chap3_model}

\begin{figure}[ht]
 \centering
	{\includegraphics[width=0.7\textwidth]{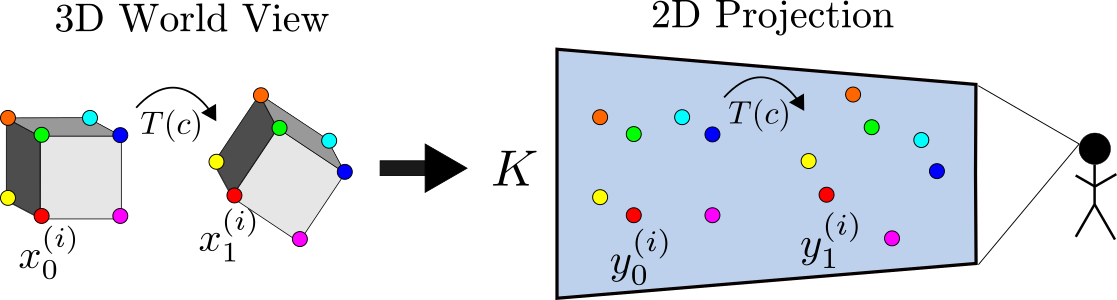}}

  \caption{\label{fig:DIVisualize}  Visualization of the 3D depth inference problem. Three dimensional points on an object are jointly transformed in the 3D world view and the visual inputs are in the form of 2D projected points.}
	
\end{figure}

We aim to develop a model for learning 3D rotational transformations from 2D projected inputs and to use that model to describe how humans may employ motion cues to recover the 3D structure of objects in their environment. This perceptual setting is visualized in Fig.~\ref{fig:DIVisualize} where an object is transforming in 3D but the visual inputs are in the form of 2D projections on the retina. In particular, each object is represented as a combination of 3D key points $\vtr{x}^{(i)} \in \mathbb{R}^3, \hspace{2mm} i = \{1,...,N_P\}$ that are projected to 2D point locations $\vtr{y}^{(i)} \in \mathbb{R}^2$. We assume rigid body motion and incorporate a transformation model that can constrain the possible 3D motion between different transformed viewpoints. This provides the structure necessary to infer depth from points on a moving object. We develop a method that uses a generative manifold model as a representation of transformations, and we show that we can learn rotational transformation operators and use them to accurately infer point depth from rotating points and scenes. We build up to the learnable model of natural transformations in two steps. In the first step, we assume the 3D rotational transformation model is known and we develop a method for inferring the depth of 2D projections of rotating points. In the second step, we utilize the depth inference approach from the first step to develop a learning model that can adapt the transformation representation to ensure it corresponds to the real world transformations.  We will preface the descriptions of each of these tasks with an overview of the transport operator model, a learnable generative manifold model.

\subsection{Transport Operator Model}\label{sec:chap3_TO}
The transport operator technique is a specific manifold learning model that learns to transform points through nonlinear Lie group operators, known as transport operators, that transverse a manifold~\citep{rao1999learning,miao2007learning,culpepper2009learning,sohl2010unsupervised,cohen2014learning,hauberg2016dreaming,connor2020representing,connor2021variational}. Lie group operators represent infinitesimal transformations which can be applied to data through an exponential mapping to transform points along a manifold. In particular, this model learns a dictionary of $M$ transport operators $\mtx{\Psi_m}$ which each represent a different transformation. These operators are effective for representing an internal transformation model for a few reasons. First, once learned, the transport operators are stored as a representation of possible transformations that may be experienced or observed. This means they can be reused in the future when the same type of transformation is visualized. Second, the transport operator model is a generative manifold model meaning it can interpolate and extrapolate new views of points undergoing a learned transformation. This provides a way of creating an internal visualization of how an object transforms similar to what humans describe when performing mental transformation tasks~\citep{zacks2005transformations}. Finally, transport operators can be used to infer the relationship between points in two separate viewpoints and define the 3D transformations between them.

With the transport operator model, the relationship between two individual 3D points $\vtr{x}^{(i)}_0$ and $\vtr{x}^{(i)}_1$ is defined as follows:
\begin{equation}
\begin{split}
\vtr{x}^{(i)}_0 = \mathrm{expm}\left(\sum_{m=1}^M{\mtx{\Psi}_m c_m}\right)\vtr{x}^{(i)}_1 + \vtr{n}, \label{eq:chap3_transOptGen} \\
\vtr{n} \sim \mathcal{N}(0,I) \quad  c_m \sim \text{Laplace}\left(0,\frac{1}{\zeta}\right),
\end{split}
\end{equation}
where $\vtr{c} \in \mathbb{R}^M$ is the set of coefficients that specifies the local structure of transformations between $\vtr{x}^{(i)}_0$ and $\vtr{x}^{(i)}_1$. Given this relationship between points, the original work from \citet{culpepper2009learning} defines the negative log posterior of the model as:
\begin{equation} \label{eq:chap3_objFun}
E_{\Psi} = \frac{1}{2}\left\|\vtr{x}^{(i)}_0 - \mathrm{expm}\left(\sum_{m=1}^M{\mtx{\Psi}_mc_m} \right) \, \vtr{x}^{(i)}_1\right\|_2^2 + \frac{\gamma}{2}\sum_m\|\mtx{\Psi}_m\|_F^2 +\zeta\|\vtr{c}\|_1,
\end{equation}
where $\|\cdot\|_F$ is the Frobenius norm and $\gamma, \zeta \geq 0$. The first term is a data fidelity term that specifies how well $\vtr{x}^{(i)}_0$ can be represented as a transformed version of $\vtr{x}^{(i)}_1$ when the transformations are constrained by the current dictionary of operators $\mtx{\Psi}$. The data fidelity objective term is an indication of how well the transport operators fit the data manifold. The second term is a Frobenius norm regularizer on the dictionary elements which constrains the growth of the dictionary magnitudes and helps identify how many operators are necessary for representing transformations on the data manifold. The third term is the sparsity regularizer which encourages each transformation between point pairs to be represented with a sparse set of coefficients.

Given a dictionary of operators $\mtx{\Psi}$, the 3D transformation between $\vtr{x}^{(i)}_0$ and $\vtr{x}^{(i)}_1$ can be estimated by inferring a set of transport operator coefficients $\vtr{c}$. This inference is performed by minimizing $E_{\Psi}$ when $\gamma = 0$. If the operators need to be learned or adapted, $E_{\Psi}$ is used as an objective for transport operator training as well. Training proceeds by alternating between coefficient inference between point pairs and gradient steps on the transport operators.

We adapt the transport operator model to use time-varying views of transforming points in a 2D projection plane to learn a generative motion model. We begin by developing an inference procedure that enables joint depth estimation and coefficient inference from pairs of 2D inputs points in different viewing frames.

\subsection{Depth Inference with Projected Inputs}\label{sec:chap3_TO_2D}

\begin{figure}[ht]
\centering
	\begin{subfigure}[b]{0.4\textwidth}
  \centering
	{\includegraphics[width=0.5\textwidth]{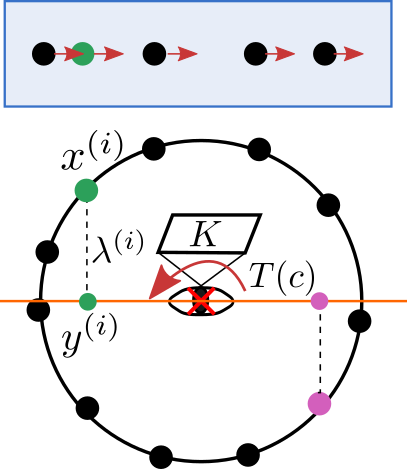}}
  \caption{}
	\label{subfig:egoSetup}
\end{subfigure}
\begin{subfigure}[b]{0.4\textwidth}
 \centering
	{\includegraphics[width=0.5\textwidth]{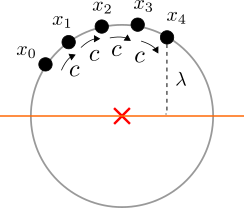}}
 \caption{}
	\label{fig:inferVis}
\end{subfigure}

  \caption{\label{fig:topVisualize} (a) Top-down views of the depth inference problem setup for points rotating on a cylinder. The 3D points $\vtr{x}^{(i)}$ have an associated depth $\lambda^{(i)}$.  Each point is projected onto the orange viewing plane using the projection matrix $\mtx{K}$. This results in the 2D projected points $\vtr{y}^{(i)}$. The 3D points are rotating counter-clockwise around the axis and the points in the blue shaded box on top indicate the direction of motion of the projected points. (b) Visualization of the inference window sequence for a single point. The inference window is made up of several frames of transformed points. We assume that the transformation speed is constant between the frames, resulting a constant coefficient value representing the transformation from one frame to the next. The depth $\lambda$ is inferred for the final frame in the sequence.}
	
\end{figure}

In this section, we will assume that the rotational transport operators are either known \emph{a priori} or already learned. We describe the training procedure in Section \ref{sec:chap3_TO_learn}. Fig.~\ref{subfig:egoSetup} shows a top-down view of the setup of this problem. The eye located at the red `x' in the center represents the viewer at the origin. The placement of the viewer at the origin is natural for learning a representation of self-motion in an egocentric viewing framework where the human is the origin. However, the model is flexible and the same set up can be used to infer object motion in the allocentric viewing framework (see Appendix~\ref{appsec:egoAllo} for more details). Each 3D point $\vtr{x}^{(i)}$ is projected onto the viewing plane to a corresponding 2D point $\vtr{y}^{(i)}$ with an associated depth $\lambda^{(i)}$: $\vtr{y}^{(i)} = \mtx{K}\vtr{x}^{(i)}$. The matrix $\mtx{K}$ is the orthographic projection matrix which is defined as $\mtx{K} = \begin{bmatrix}1&0&0\\0&1&0\end{bmatrix}$ in all of our experiments. This projection matrix corresponds to setting the viewing plane to the  $xy$-plane and defining the unknown depth as the $z$-coordinate of the 3D input points. It is assumed that the $\mtx{K}$ is known during processing. We observe $N_P$ points transforming jointly on a rigid object and concatenate all $N_P$ points into a matrix: $\mtx{X}_0 = \begin{bmatrix}\vtr{x}_0^{(1)} .... \vtr{x}_0^{(N_P)}\end{bmatrix}$.

The relationship between points in two consecutive frames at $t=0$ and $t=1$ is defined as:
\begin{equation}\label{eq:y0_eq_multi}
    \mtx{Y}_0 = \mtx{K}\mtx{T}(\vtr{c})\widehat{\mtx{X}}_1\left(\vtr{\lambda}\right) + \mtx{W},
\end{equation}
where $\mtx{W}$ is a Gaussian noise matrix, $\widehat{\mtx{X}}_1$ is a matrix of estimated 3D point locations associated with $\mtx{Y}_1$ and $\mtx{T}(\vtr{c})$ is the matrix exponential of a weighted combination of transport operators which can each represent a different type of motion:
\begin{equation}
    \mtx{T}(\vtr{c}) = \mathrm{expm}\left(\sum_{m=1}^M{\mtx{\Psi}_m c_m}\right).
\end{equation}

In \eqref{eq:y0_eq_multi} we define $\mtx{Y}_0$ at $t=0$ as a transformation of points at $t=1$ in order to estimate the 3D point locations in $\widehat{\mtx{X}}_1$ at $t=1$ in a causal manner as we describe below. To compute $\widehat{\mtx{X}}_1$, we reverse the process of the projection matrix in two steps.  First, we concatenate the $\mtx{Y}_1$ with a vector of zeros in the $z$-coordinate position that is lost during projection:
\begin{equation}
    \mtx{\widetilde{X}}_1 = \begin{bmatrix}\vtr{y}_1^{(1)} & \cdots & \vtr{y}_1^{(N_P)}\\0 & \cdots & 0\end{bmatrix}.
\end{equation}

Second, we add the depths to the newly introduced dimension. To do this, we compute the outer product between the standard basis vector associated with the axis lost during projection $\vtr{e}_{z}$ and the depth vector $\lambda \in \mathbb{R}^{N_P}$, resulting in a matrix with two rows of zeros and one row containing the estimated depths, and add that to $\mtx{\widetilde{X}}_1$:
\begin{equation}
    \widehat{\mtx{X}}_1\left(\vtr{\lambda}\right) = \mtx{\widetilde{X}}_1 + \vtr{e}_z \lambda^\top.
\end{equation}

This model for incorporating the estimated depths can be integrated into the data fidelity term of the objective function in \eqref{eq:chap3_objFun} and used for jointly inferring the depth $\mtx{\lambda}$ and coefficients $\vtr{c}$ between point pairs. 
\begin{align}
    L_{\text{df}} &= \frac{1}{2}\sum_{i=1}^{N_P}\|\vtr{y}^{(i)}_0 - \mtx{K}\mtx{T}(\vtr{c})\widehat{\vtr{x}}^{(i)}_1\left(\vtr{\lambda}^{(i)}\right)\|_2^2  \\
    &= \frac{1}{2}\text{trace}\left(\left(\mtx{Y}_{0} - \mtx{K}\mtx{T}(\vtr{c})\widehat{\mtx{X}}_{1}(\vtr{\lambda})\right)^\top\left(\mtx{Y}_{0} - \mtx{K}\mtx{T}(\vtr{c})\widehat{\mtx{X}}_{1}(\vtr{\lambda})\right)\right).
\end{align}

We add two additional constraints to this model to improve the consistency of accurate depth estimation. First, we incorporate a Gaussian prior on the depths which constrains them to magnitudes consistent with the ground truth depths of the rotating objects. 
Second, we group several consecutive views of the transforming points to reverse the projection procedure on points in the final frame in the sequence. We refer to this sequence of frames as the inference window. During inference and learning, we use ground truth knowledge of point correspondences between frames. From this inference window, we can obtain a causal estimate of the depth in the final frame and infer a fixed set of coefficients that represents the transformations between each consecutive view. This assumes a fixed transformation speed over multiple frames which can be seen as an extension of the slowness principle to natural transformations that persist over time~\citep{wiskott2002slow}. Fig.~\ref{fig:inferVis} shows this setting where the same coefficients $\vtr{c}$ are inferred between points in each neighboring frame and the depth is inferred for the final point in the sequence. Using more than two motion frames for depth inference provides additional information that can be used to resolve depth ambiguities. To model this setting, we generalize \eqref{eq:y0_eq_multi} for $N_T$ viewing frames:
\begin{equation}\label{eq:yN_step}
    \mtx{Y}_{N_T-n} = \mtx{K}\mtx{T}^n(\vtr{c})\widehat{\mtx{X}}_{N_T}\left(\vtr{\lambda}\right) =  \mtx{K}\mtx{T}(n\vtr{c})\widehat{\mtx{X}}_{N_T}\left(\vtr{\lambda}\right), \quad n = \{1,...,N_T\},
\end{equation}
where the change from $\mtx{T}^n(\vtr{c})$ to $\mtx{T}(n\vtr{c})$ is possible because raising an exponent to the power of $n$ is the same as applying the same transformation $\mtx{T}(\vtr{c})$ $n$ times and thus multiplying its transformation coefficients by $n$.

We define an objective that leverages multiple views and a depth regularizer: 
\begin{equation}\label{eq:DI_obj}
\begin{split}
    L = \frac{1}{2N_T}\sum_{n=1}^{N_T}\sum_{i=1}^{N_P}\left[\|\vtr{y}^{(i)}_{N_T-n} - \mtx{K}\mtx{T}(n\vtr{c})\widehat{\vtr{x}}^{(i)}_{N_T}(\vtr{\lambda})\|_2^2\right] + \zeta\|\vtr{c}\|_1 \\ + \frac{\beta}{2}\|\vtr{\lambda}\|^2_2 + \frac{\gamma}{2}\sum\limits_m\|\mtx{\Psi}_m\|_F^2.
    \end{split}
\end{equation}

With this objective, the depth $\lambda$ and the coefficients $\vtr{c}$ can be jointly inferred for sequences of transforming points. See Section \ref{appsec:DIInferDet} for more details on the inference process.

To highlight the effectiveness of this inference model, we will examine how accurately it can infer depths and transformations using ground truth rotational operators and explore the requirements for the inputs that lead to robust depth estimation.

\section{Results}
\subsection{Depth Inference Experiments}\label{sec:chap3_DIExp}

\begin{figure}[t]

\centering
\begin{subfigure}[b]{0.31\textwidth}
  \centering
	{\includegraphics[width=0.98\textwidth]{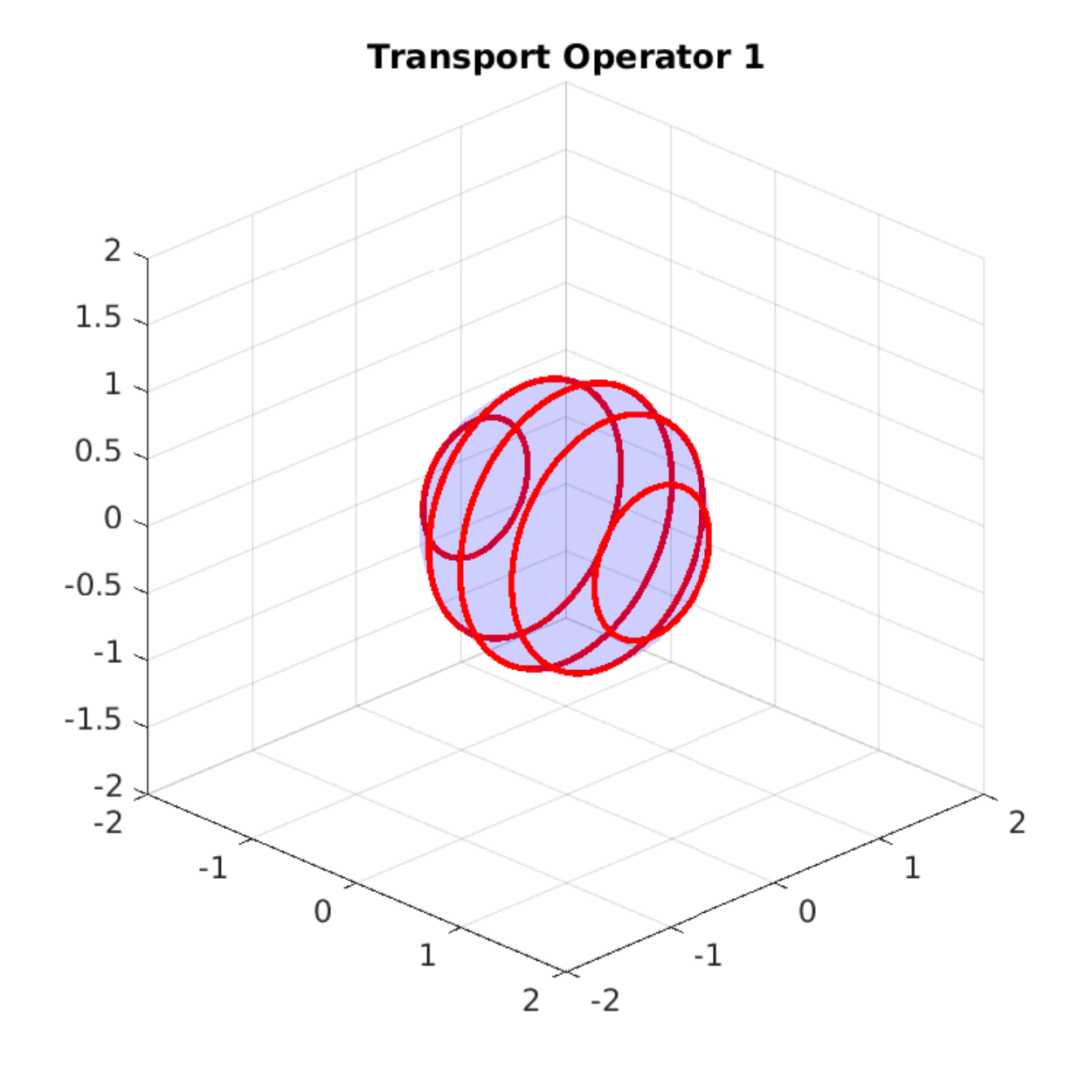}}
  \caption{}
	\label{subfig:gtrotOpt_1}
\end{subfigure}
\begin{subfigure}[b]{0.31\textwidth}
 \centering
	{\includegraphics[width=0.98\textwidth]{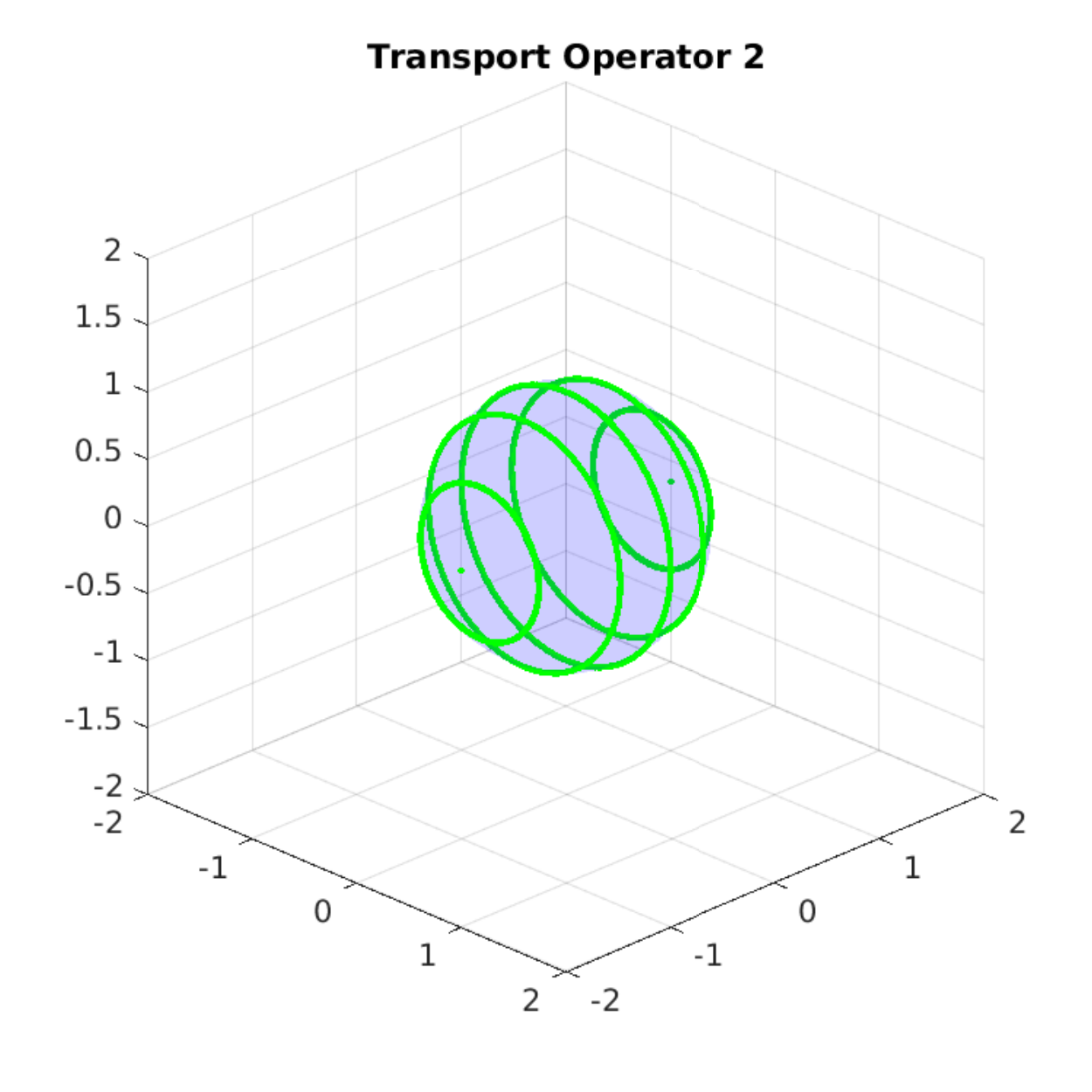}}
 \caption{}
	\label{subfig:gtrotOpt_2}
\end{subfigure}
\begin{subfigure}[b]{0.31\textwidth}
 \centering
	{\includegraphics[width=0.98\textwidth]{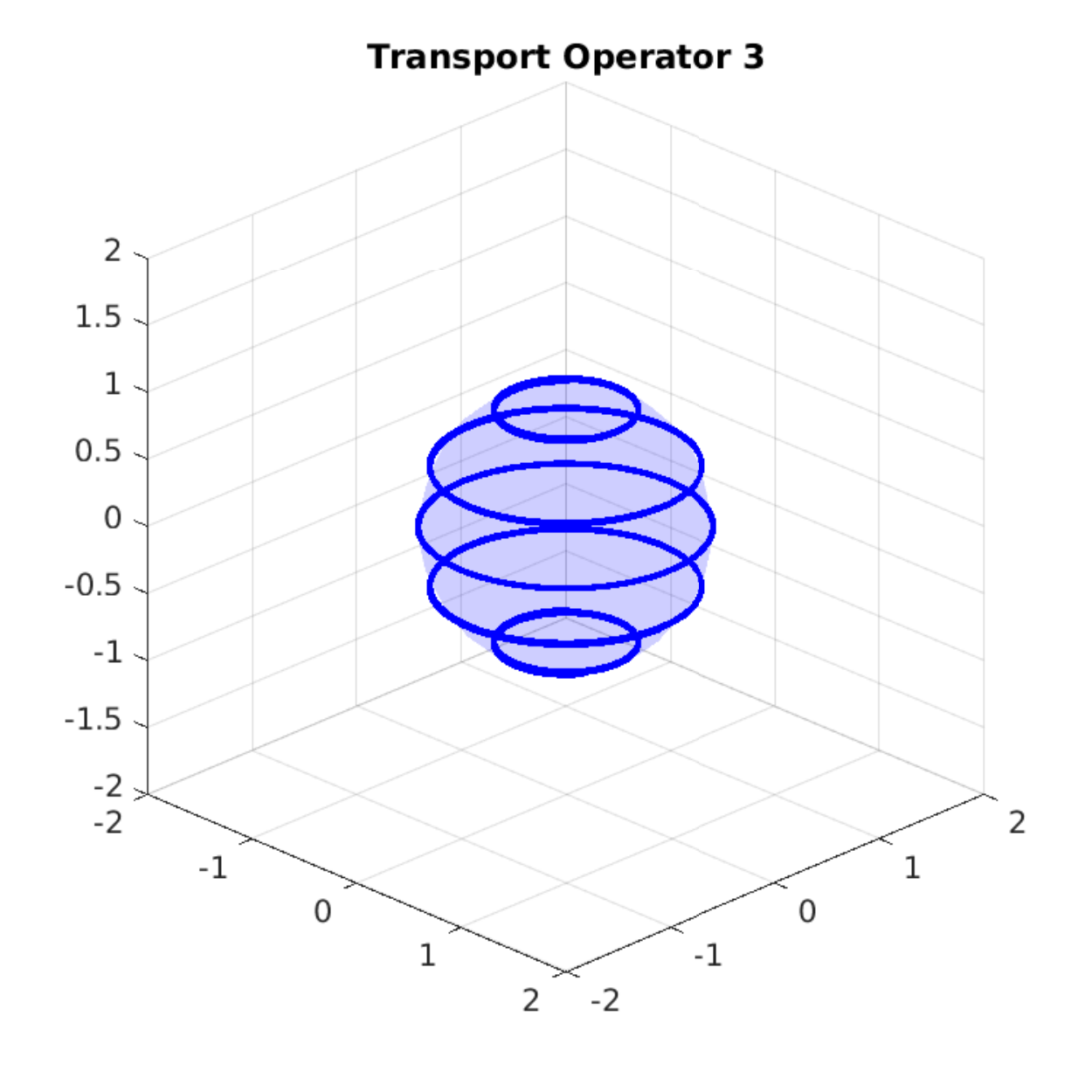}}
 \caption{}
	\label{subfig:gtrotOpt_3}
\end{subfigure}

  \caption[Trajectories generated by ground truth rotational transport operators.]{\label{fig:gtrotOpt} Trajectories generated by ground truth rotational transport operators. Each line represents the trajectory of an individual transport operator dictionary element applied to one of several example starting points selected on the sphere. These three operators generate rotation around each of the three principle axes.}
	
\end{figure}

Three-dimensional rotational matrices can be defined as elements of the 3D rotational group SO(3) and ground truth rotational transport operators can be derived from elements of the $\mathfrak{so}(3)$ Lie algebra~\citep{hall2015lie}. Fig.~\ref{fig:gtrotOpt} shows the trajectories of these ground truth 3D rotational operators, each rotating around one of the principal axes. These plots are generated by selecting a few example starting points on a sphere and applying individual dictionary elements $\mtx{\Psi_m}$ to each point as they evolve over time: $\vtr{x}^{(i)}_t = \mathrm{expm}(\mtx{\Psi}_m \frac{t}{T})\vtr{x}^{(i)}_0$, $t = 0,...,T$.

Using \eqref{eq:DI_obj} and the fixed $\mtx{\Psi}$ representing ground truth rotational operators, we jointly infer the coefficients and depths from a sequence of transforming points. Fig.~\ref{fig:inferVis} shows a visualization of this inference setting for a single point. Given $N_T$ views of rotating points, we infer the depth $\lambda$ for the projected points in the last viewpoint $\mtx{Y}_{N_T}$ as well as the coefficients $\vtr{c}$ that correspond to the shared transformation between every pair of consecutive views in the sequence. This ensures that the depths at time $N_T$ are inferred by samples preceding it in the motion sequence, making this a causal estimate.  Because the objective \eqref{eq:DI_obj} is nonconvex, inference may result in local minima. To avoid the local minima, we perform inference for the same inference window several times using several random restarts. That is, we randomly sample a new intialization and infer coefficients and depths using that starting point. This often results in different final inferred outputs. We choose the inferred output associated with the lowest final objective from inference.

\begin{figure}[t]

\centering
\begin{subfigure}[b]{0.22\textwidth}
  \centering
	{\includegraphics[width=0.96\textwidth]{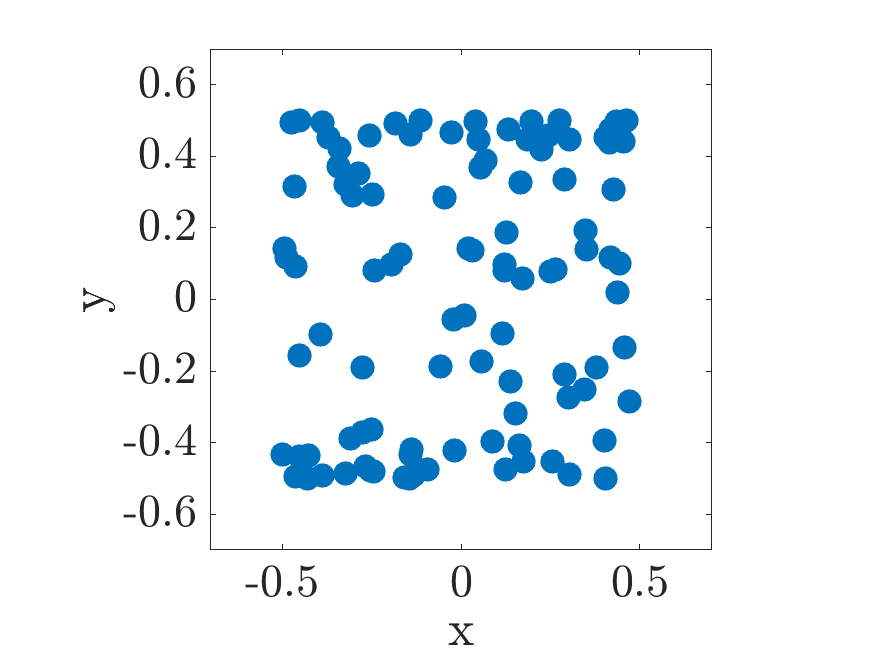}}
  \caption{}
	\label{subfig:projCyl}
\end{subfigure}
\begin{subfigure}[b]{0.22\textwidth}
 \centering
	{\includegraphics[width=0.96\textwidth]{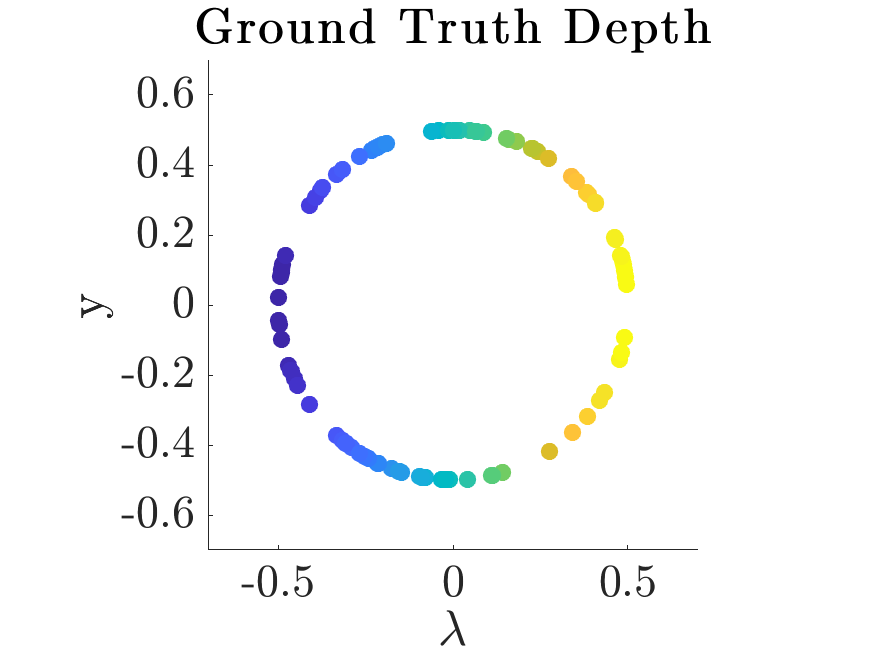}}
 \caption{}
	\label{subfig:depthCylGT}
\end{subfigure}
\begin{subfigure}[b]{0.22\textwidth}
 \centering
	{\includegraphics[width=0.96\textwidth]{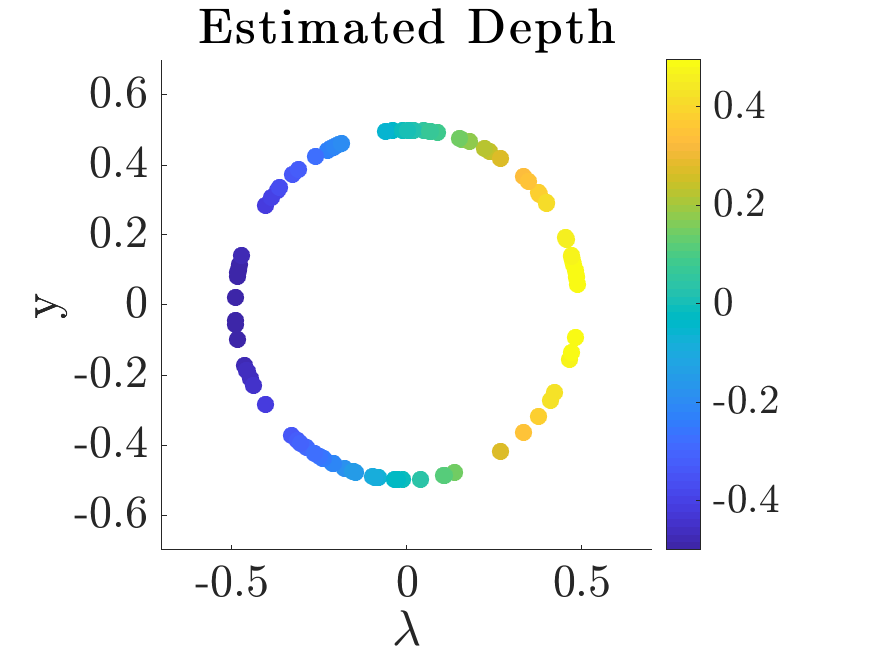}}
 \caption{}
	\label{subfig:depthCylEst}
\end{subfigure}

\begin{subfigure}[b]{0.22\textwidth}
  \centering
	{\includegraphics[width=0.96\textwidth]{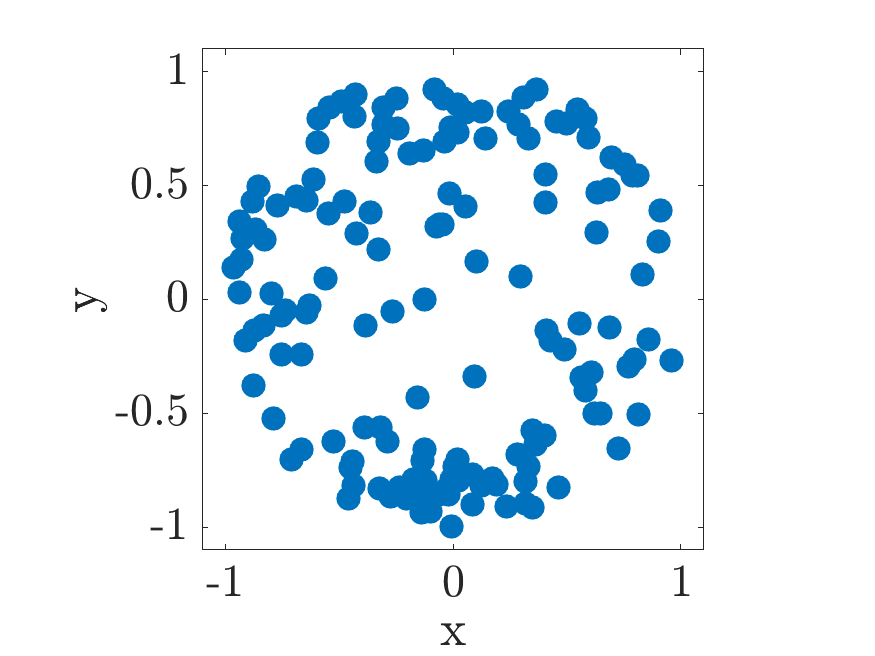}}
  \caption{}
	\label{subfig:projSph}
\end{subfigure}
\begin{subfigure}[b]{0.22\textwidth}
 \centering
	{\includegraphics[width=0.96\textwidth]{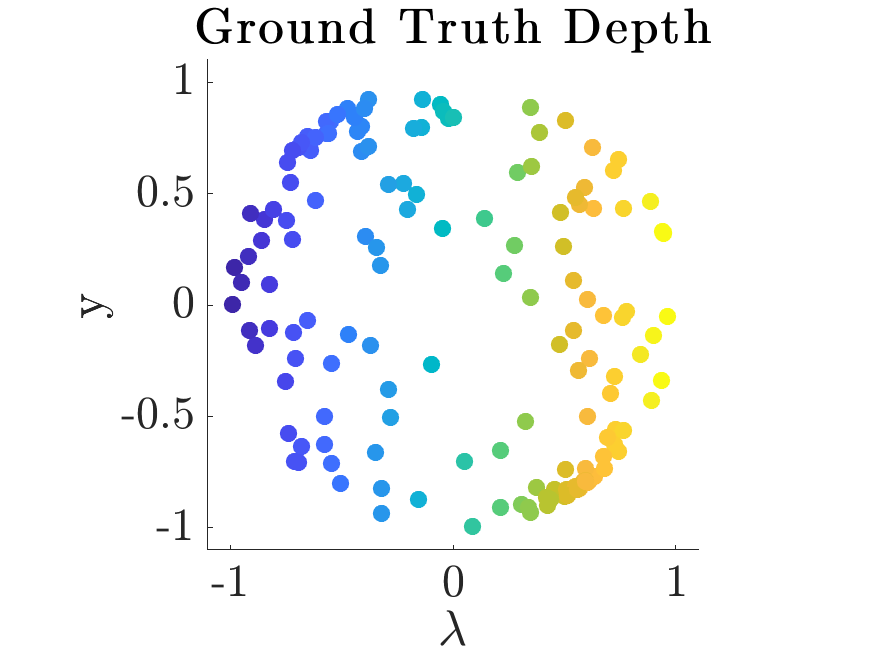}}
 \caption{}
	\label{subfig:depthSphGT}
\end{subfigure}
\begin{subfigure}[b]{0.22\textwidth}
 \centering
	{\includegraphics[width=0.96\textwidth]{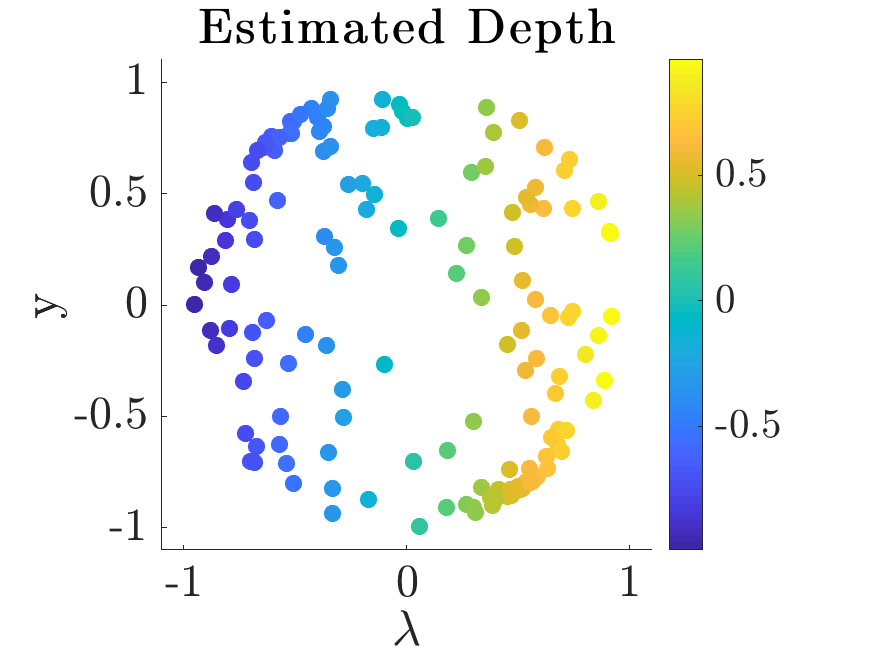}}
 \caption{}
	\label{subfig:depthSphEst}
\end{subfigure}

\begin{subfigure}[b]{0.22\textwidth}
  \centering
	{\includegraphics[width=0.96\textwidth]{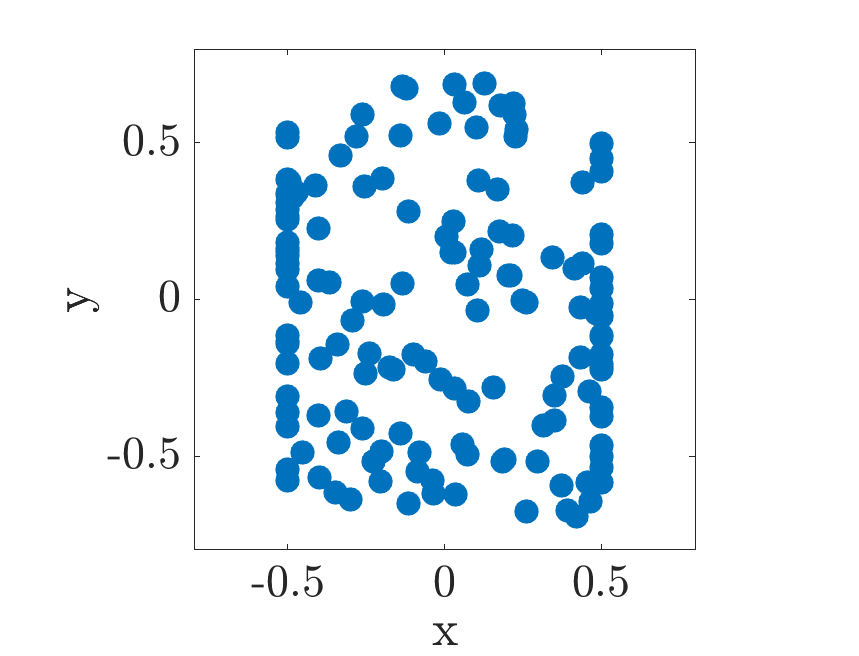}}
  \caption{}
	\label{subfig:projCube}
\end{subfigure}
\begin{subfigure}[b]{0.22\textwidth}
 \centering
	{\includegraphics[width=0.96\textwidth]{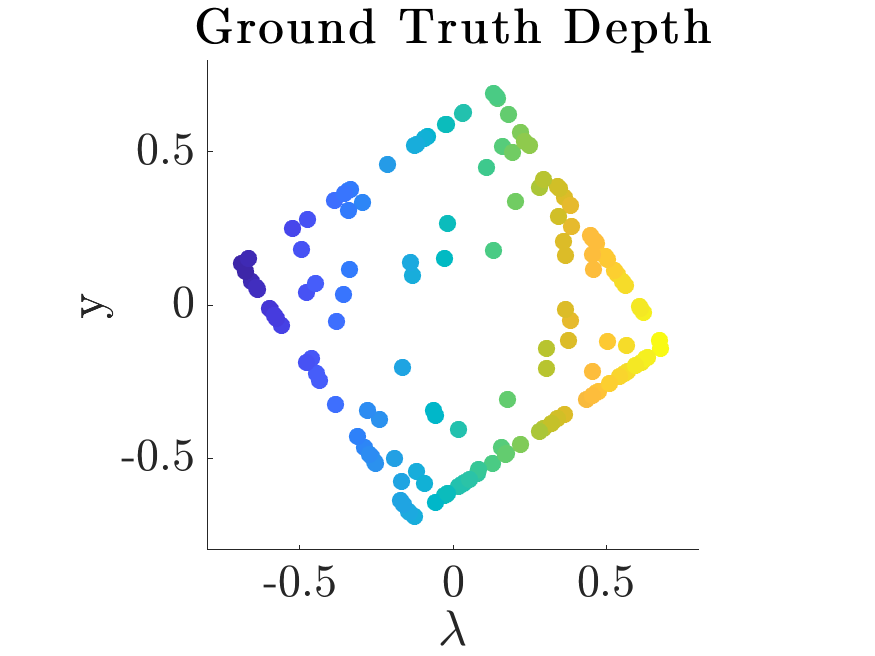}}
 \caption{}
	\label{subfig:depthCubeGT}
\end{subfigure}
\begin{subfigure}[b]{0.22\textwidth}
 \centering
	{\includegraphics[width=0.96\textwidth]{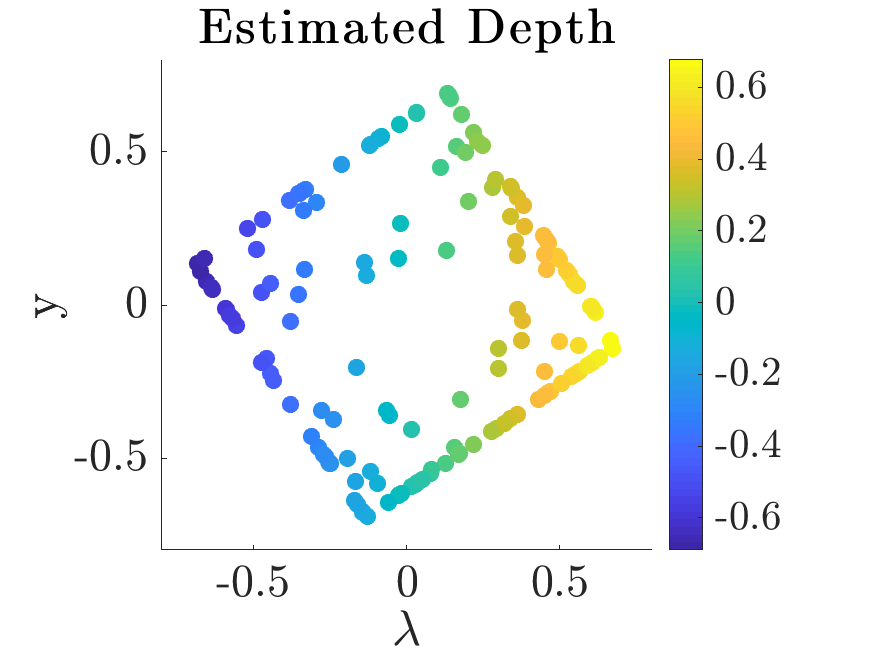}}
 \caption{}
	\label{subfig:depthCubeEst}
\end{subfigure}

  \caption[Examples of inferred depths for points on the surface of shapes.]{\label{fig:inferEx} Inferred depths for points on the surface of different shapes. The first column shows the 2D point projections in the final viewing plane. The second column shows a side view of the ground truth 3D point stimuli where the $x$-axis in the plots is the depth axis. The third column shows a side view of the estimated depths for the projected points. The points in the second and third columns are colored by the ground truth depth.  (a-c) Cylinder (d-f) Sphere (g-i) Cube.}
	
\end{figure}

Fig.~\ref{fig:inferEx} shows examples of depth inference for points on the surfaces of three different shapes. The plots in the first column show the visual input of projected points in the final viewing plane of the sequence. The plots in the second column show a side view of the point stimuli where the ground truth depth locations are shown on the $x$-axis of the plot. The plots in the third column show a side view with the estimated depth locations for each of the points. In both the second and third columns, the points are colored by the ground truth depths. This shows that the estimated depths correspond with the ground truth depths for a variety of shapes.

We quantitatively evaluate the accuracy of the inferred depths for many trials to analyze the impact of various model parameters. There are three parameters of particular interest during inference. First we are interested in the impact of the perceptual extent of rotation viewed in a sequence of frames. The perceptual extent of rotation is a combination of two parameters, the number of frames in a rotation sequence $N_T$ and the ground truth rotation angle between each frame in the rotation sequence $\theta$. The full angular extent of rotation viewed is $\theta_{\text{path}} = N_T \theta$. Experiments have shown that larger angular extents of rotation can lead to more accurate depth estimates for human subjects~\citep{hildreth1990perceptual}. Next we are interested in the impact of the number of jointly transforming points $N_P$. This indicates the amount of coherent rotational motion viewed in the input stimulus. Psychophysical experiments have been run that indicate a greater number of coherently moving points leads to a more robust depth percept~\citep{todd1988apparent,dosher1989ratings,sperling1989kinetic,braunstein1987minimum}. 

In order to quantitatively evaluate the success of inferred depth, we use two metrics. The first is the mean squared error between the estimated depths and the ground truth depths for the $N_P$ rotating points. Ideally depth inference would result in low MSE between the estimated depth and the ground truth depth. However, with a rotational transformation, as we are working with here, there exists a depth-angle ambiguity. Namely, when viewing projected points $\vtr{y}_0^{(i)}$ and $\vtr{y}_1^{(i)}$ from two separate views, they could be either projections of points with large depths that undergo rotation with a smaller angle or points with small depths that undergo rotation with a larger angle (see Appendix~\ref{appsec:depthAng} for a visualization). While it is not ideal for the depth to be off by a scaling factor, the inferred structure can still be accurate. Additionally, this depth ambiguity is observed in experiments with human subjects~\citep{todd1988apparent}. In psychophysical experiments, one metric for determining accuracy of a percept is comparing the estimated ordering of point depths to the ground truth ordering of point depths~\citep{hildreth1990perceptual}. To analyze the accuracy of the inferred structure in the presence of potential scaling in depth, we compare the ordering of the inferred depths to that of the ground truth depths using the Kendall's Tau rank correlation coefficient~\citep{kendall1938new}. We compare the Kendall's Tau between all points $N_P$ as well as between five randomly selected points. We choose to compare the ordering of five randomly selected points in order to define a metric that can be used to fairly compare the performance as the number of points increases. With greater $N_P$, even if depths are accurate within some error range, there is a greater chance of incorrectly ordering a few points because there is a greater point density. Therefore, comparing five randomly sampled points should provide a consistent metric as we vary $N_P$.  

\begin{figure}[t]

\centering
\begin{subfigure}[b]{0.4\textwidth}
  \centering
	{\includegraphics[width=0.94\textwidth]{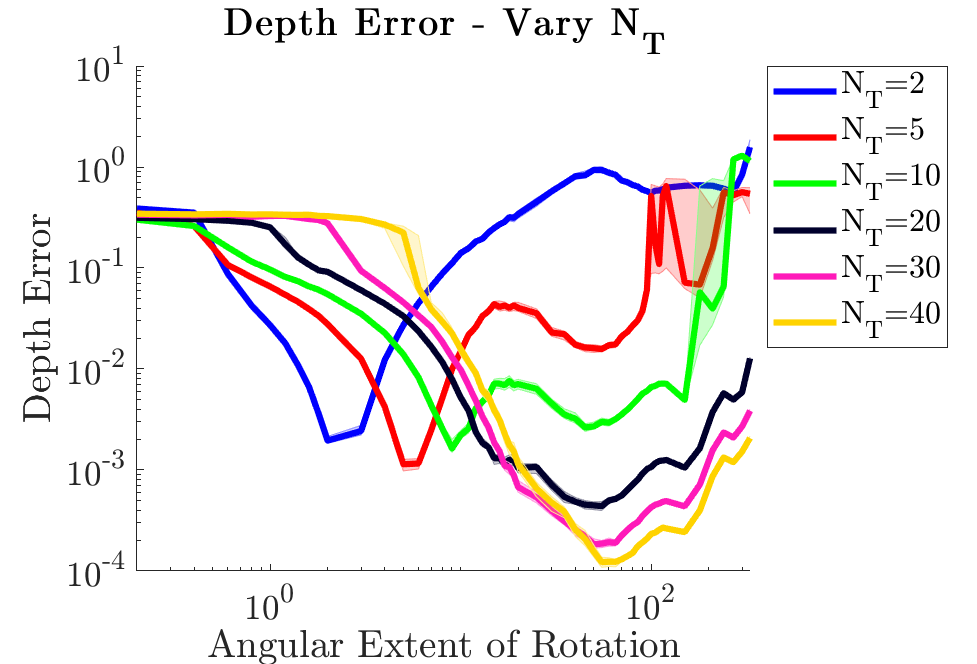}}
  \caption{}
	\label{subfig:depthErr_rotExt_Nt}
\end{subfigure}
\begin{subfigure}[b]{0.4\textwidth}
 \centering
	{\includegraphics[width=0.94\textwidth]{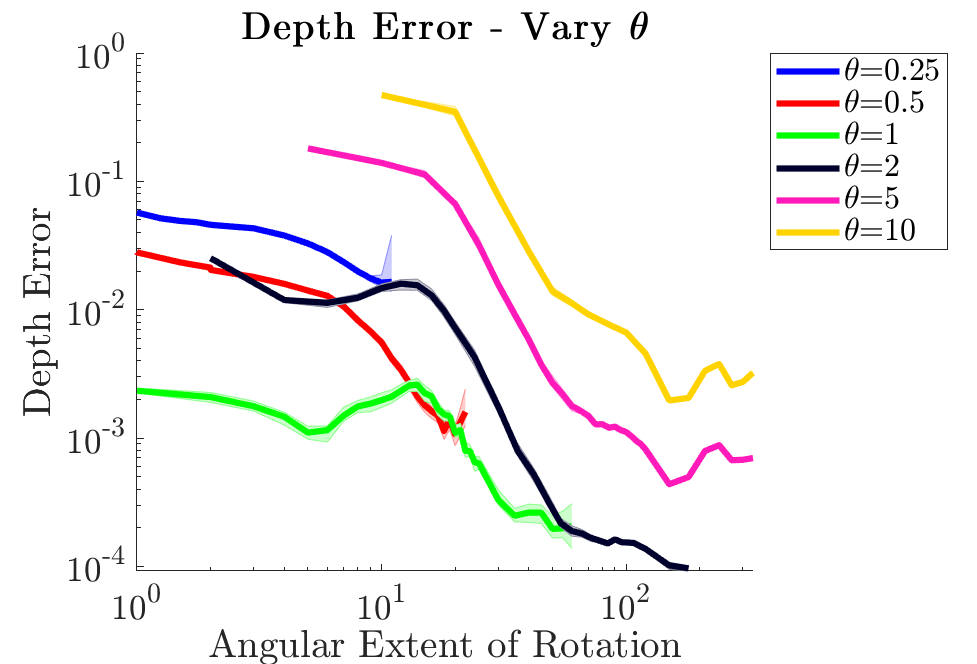}}
 \caption{}
	\label{subfig:depthErr_rotExt_deg}
\end{subfigure}
\begin{subfigure}[b]{0.24\textwidth}
  \centering
	{\includegraphics[width=0.98\textwidth]{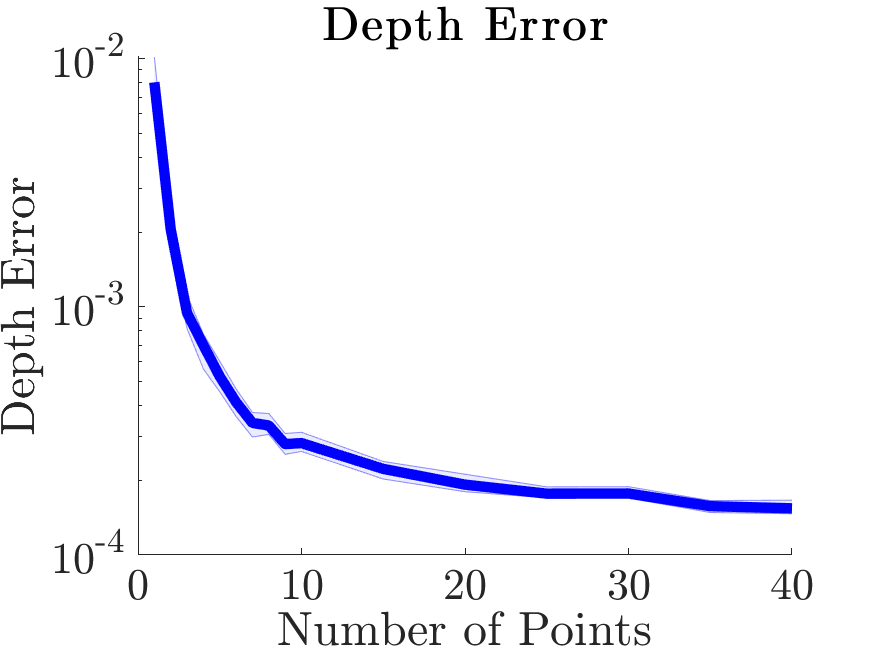}}
  \caption{}
	\label{subfig:depthErr_numPts}
\end{subfigure}
\begin{subfigure}[b]{0.24\textwidth}
 \centering
	{\includegraphics[width=0.98\textwidth]{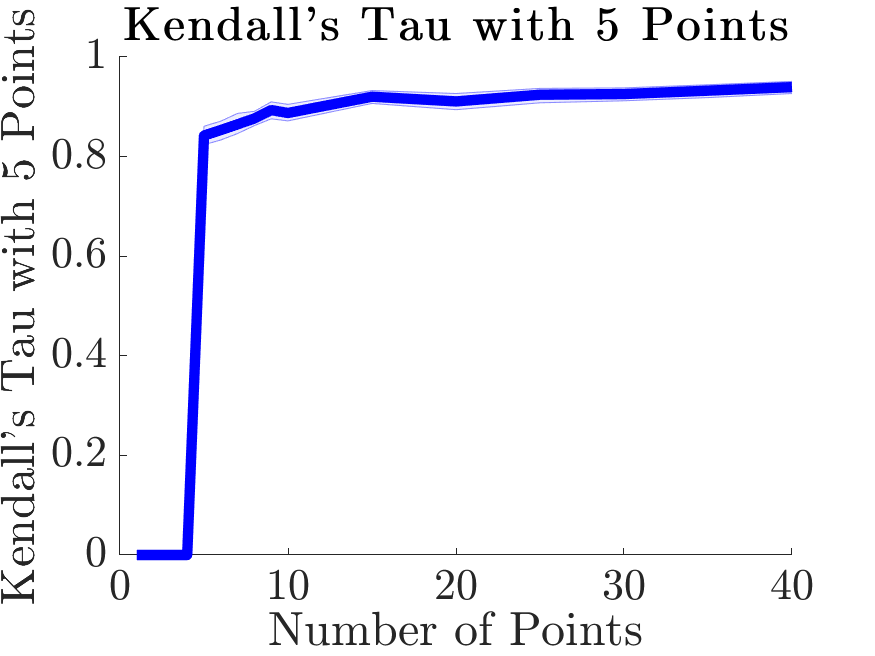}}
 \caption{}
	\label{subfig:kt5_numPts}
\end{subfigure}
\begin{subfigure}[b]{0.24\textwidth}
  \centering
	{\includegraphics[width=0.98\textwidth]{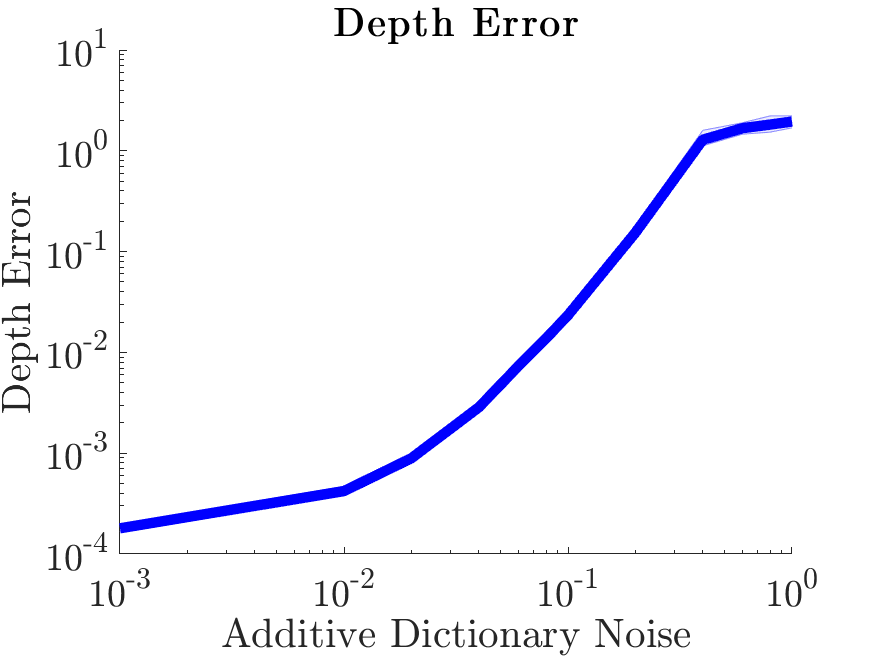}}
  \caption{}
	\label{subfig:depthErr_psiN}
\end{subfigure}
\begin{subfigure}[b]{0.24\textwidth}
 \centering
	{\includegraphics[width=0.98\textwidth]{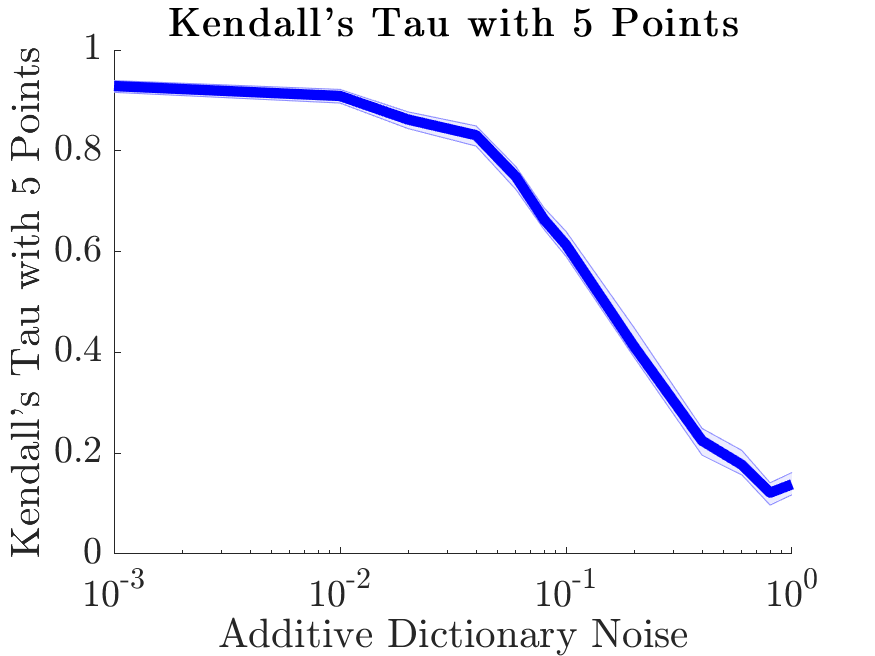}}
 \caption{}
	\label{subfig:kt5_psiN}
\end{subfigure}

  \caption{\label{fig:quant_rotExt} Quantitative metrics for depth inference when varying angular extent of rotation, the number of coherently transforming points $N_P$, and the standard deviation of Gaussian noise added to the ground truth operators. (a) Median depth error as angular extent increases. Each line is generated with different numbers of frames in the inference window $N_T$. The optimal performance occurs for angular extents in the range of $N_T$ to $2N_T$. (b) Median depth error as angular extent increases. Each line is generated with different angles of rotation between sequence frames $\theta$. A rotation angle of $\theta = 2$ results in the lowest depth error at $180 \degree$ of rotation. (c) Median depth error as $N_P$ increases. (d) Mean Kendall's Tau for 5 randomly selected points as $N_P$ increases. Values of this metric for $N_P < 5$ are set to zero because there are not enough points to compare five randomly selected points.  (e) Median depth error as the standard deviation of dictionary noise increases. (f) Mean Kendall's Tau for 5 randomly selected points.} 
	
\end{figure}

Fig.~\ref{subfig:depthErr_rotExt_Nt} and \ref{subfig:depthErr_rotExt_deg} show the median depth error as we vary the angular extent of rotation $\theta_{\text{path}}$. In Fig.~\ref{subfig:depthErr_rotExt_Nt}, each line represents a different number of frames $N_T$. The error bars in all plots represent the bootstrap confidence interval.  For each line in Fig.~\ref{subfig:depthErr_rotExt_Nt}, because the values of $N_T$ are fixed, moving along the $x$-axis corresponds to increasing the angle between frames $\theta$. Each of these lines has a clear minimum and this minimum occurs at an angular extent in the range of $N_T$ to $2N_T$. This corresponds to an angle $\theta$ between frames of $1\degree-2\degree$. Up to this minimum value, the depth error decreases as the rotational extent increases, indicating the benefit of a more complete view of the rotational sequence. For smaller rotational extents, the inference results in low objective function values but large depth errors. This indicates that there are local minima of the optimization that achieve low objective function values but do not accurately reflect the rotation sequence and point depth. With greater rotational extents, the low objective function values correspond more directly to accurate depth inference. 

As $\theta$ increases, the magnitudes of the coefficients corresponding to the true rotation increase. This presents a challenge for the non-convex objective because the space of possible coefficient values increases. Therefore, randomly selecting initializations of the coefficients in the neighborhood of the true minimum is less likely which results in more solutions that correspond to inaccurate local minima. The optimal values between $N_T$ and $2N_T$ have a large enough rotational extent such that low objective function values correspond to accurate depths but small enough rotational extent that we can randomly initialize inference with coefficients that result in minima close to the ground truth depth values. Fig.~\ref{subfig:depthErr_rotExt_deg} breaks down the performance for individual values of $\theta$. As $\theta$ increases up to $\theta = 2\degree$, the depth error decreases with increasing $\theta$. For $\theta$ greater than $5\degree$, the performance starts to degrade. This is consistent with the patterns in Fig.~\ref{subfig:depthErr_rotExt_Nt}, and it indicates that rotation angles that are too large between frames (corresponding to fast rotational motion) results in less accurate depth inference\footnote{We should note that inference optimization experiences inaccuracy even with a large number of random restarts as greater number of frames are used in the inference window, leading to large increases in depth error for $N_T > 50$ in many settings. Therefore, in Fig.~\ref{subfig:depthErr_rotExt_deg}, we only display lines until an angular extent of rotation for which the non-convex optimization inaccuracy impacts the solutions. The angular extent where this occurs is smaller for the smaller values of $\theta$ because they require larger values of $N_T$ to achieve the same angular extent of rotation.}. In the remaining tests of inference performance with ground truth operators, unless otherwise stated, we set $N_T = 30$ and $\theta = 2$. See \ref{appsec:DIInferDet} for more details on model parameters. 

Fig.~\ref{subfig:depthErr_numPts} and \ref{subfig:kt5_numPts} show the median depth error and the mean Kendall's Tau as we vary $N_P$. This shows that the depth estimation improves as more points are added with a large performance improvement from $N_P = 1$ to $N_P = 10$. We reason that this improvement is due the reduction in transformation ambiguity that results from seeing more points rotating jointly. The greater number of points on a rigid object undergoing the same transformation, the more information our model has about the accurate transformation and depth. Going forward, we use $N_P = 20$. Research in structure from motion has shown that increasing the number of transforming points improves the general depth percept~\citep{todd1988apparent,dosher1989ratings,sperling1989kinetic} but it may not increase the accuracy of the inferred depths~\citep{braunstein1987minimum}.

The final quantity we analyze in this controlled setting with ground truth rotational operators is the effect of adding Gaussian noise to the operators. Noisy operators depart from the ground truth rotational transformations and analyzing the performance with noisy operators can indicate the impact of accurate rotational transformation models on effective depth inference. Fig.~\ref{subfig:depthErr_psiN} and \ref{subfig:kt5_psiN} show the median depth error and mean Kendall's Tau metrics as noise is added to the ground truth operators. Both metrics indicate that the depth inference is robust to noise with a standard deviation of around $10^{-3}$ - $10^{-2}$ but performance decreases sharply with noise larger than that. This shows that the model can perform effectively with some transformation inaccuracy but performance decreases with increasingly inaccurate transport operators. This highlights the necessity of accurate rotational operators and inspires the learning and adaptation procedure introduced and analyzed in the next section.

\subsection{Learning 3D Transport Operators from 2D Projected Inputs}\label{sec:chap3_TO_learn}

A stated goal of this work is to develop a model that can learn 3D transformational representations from rotating 2D projected input points. The learning procedure is a straight-forward extension of the coefficient and depth inference model from the previous sections. Training of the transport operator dictionary elements is performed using gradient descent. For each training step, a sequence of projected rotated points $\mtx{Y}_n, \hspace{2mm} n = \{1,...,N_T\}$ is generated. First the dictionary weights are fixed and the depth and coefficients are inferred. Then, fixing the depth and coefficients, the gradient on the dictionary elements is computed using the objective in \eqref{eq:DI_obj} with $\zeta = \beta = 0$. If this gradient step improves the objective, then it is accepted. Otherwise, the step is rejected and the learning rate is decreased. See Section~\ref{appsec:DITrainDet} for more details on the training procedure.

\begin{figure}[ht]

  \centering
	{\includegraphics[width=0.97\textwidth]{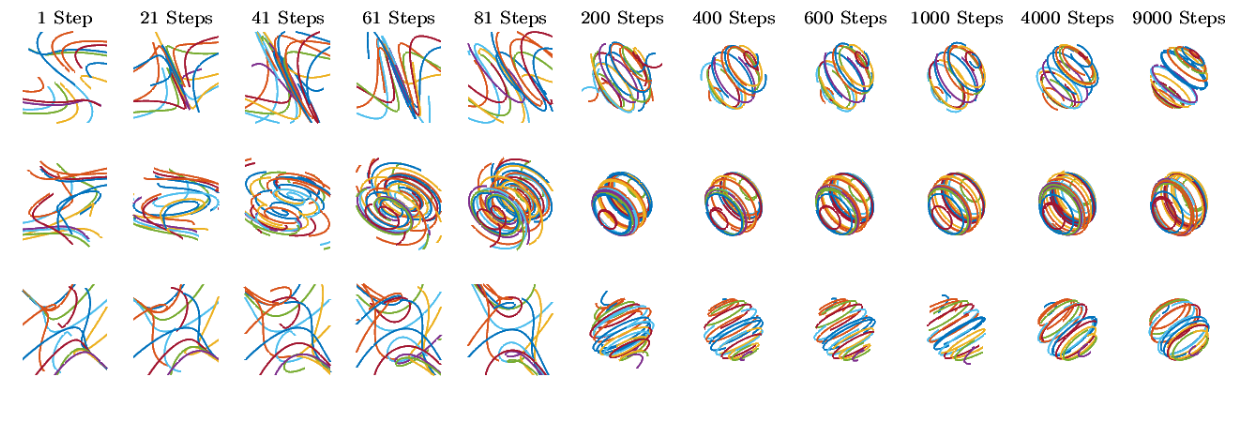}}

  \caption[Transport operator trajectories during training.]{\label{fig:trainTraj} Transport operator trajectories during training. Each row represents one of the three learned operators. Each column shows the trajectories at a different training step. The operators begin with random initializations at step 1 and quickly reach a rotation structure around 200 steps. From 200 steps to 9000 steps, the operators vary relatively slowly, resulting in operators with clear rotational structure at the end of training.}
	
\end{figure}

With this training procedure, we are able to learn rotational transport operators from randomly initialized operators. Fig.~\ref{fig:trainTraj} shows the trajectories of the operators during one training run in which the number of dictionary elements $M$ is set to 3. At the beginning of training, the trajectories do not correspond to common geometric transformations but they quickly adapt to represent near-rotational operators with trajectories similar to the ground truth operators shown in Fig.~\ref{fig:gtrotOpt}. In Section \ref{appsec:DITrainDet} we show an example of learning rotational operators from a dictionary with six operators. 

\begin{figure}[t]

\centering
\begin{subfigure}[b]{0.48\textwidth}
  \centering
	{\includegraphics[width=0.8\textwidth]{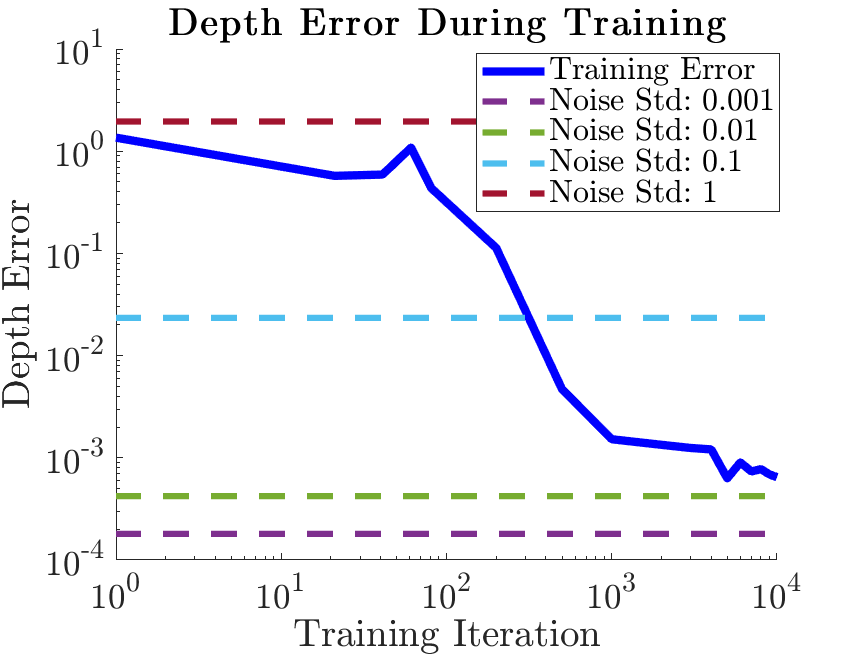}}
  \caption{}
	\label{subfig:trainDepthErr}
\end{subfigure}
\begin{subfigure}[b]{0.48\textwidth}
 \centering
	{\includegraphics[width=0.8\textwidth]{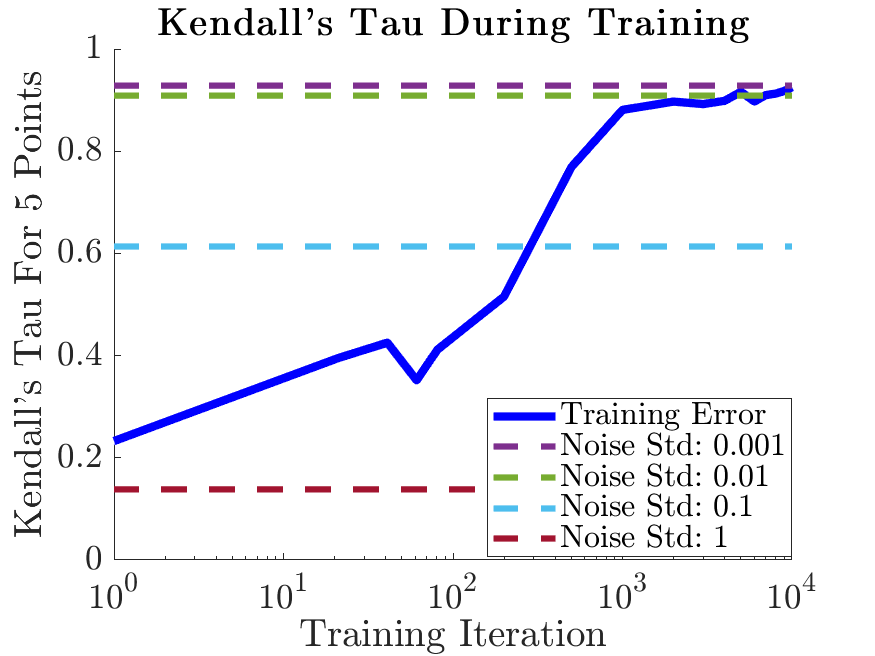}}
 \caption{}
	\label{subfig:trainkt5}
\end{subfigure}

  \caption[Inferred depth metrics using operators at different steps in  training.]{\label{fig:trainMetrics} Inferred depth metrics using operators at different steps in training. Dotted lines represent values of error metrics for depths inferred using ground truth operators with additive Gaussian noise with the standard deviation specified in the legend. These values are obtained from the plots in Fig.~\ref{subfig:depthErr_psiN} and~\ref{subfig:kt5_psiN}. (a) Median depth error for depth inference performed with operators at different steps in training. The depth error decreases significantly after 200 training steps and continues to decrease until the end of training.  The depth error achieves a value consistent with the estimates using ground truth operators with a additive Gaussian noise with a standard deviation of $10^{-2}$. (b) Mean Kendall's Tau for 5 randomly selected points. The Kendall's Tau increases significantly around 200 training steps and starts to plateau around 1000 training steps. The Kendall's Tau reaches a value consistent with ground truth operators with a noise standard deviation of $10^{-3}$. }
	
\end{figure}

We quantitatively compare these operators to the ground truth operators using the same depth MSE and Kendall's Tau metrics employed to analyze inference success. We can compute the depth error and Kendall's Tau metrics for inferred depths using operators at various points during training and compare them to the metric values resulting from depth inference using the ground truth operators with noise added. This gives us a proxy for estimating the deviation between the learned operators and ground truth rotational operators. In Fig.~\ref{fig:trainMetrics}, we show the depth error and Kendall's Tau for depths inferred using transport operators at different points in the training procedure. For reference, we also plot straight lines which correspond to the values for these metrics in Fig.~\ref{subfig:depthErr_psiN} and \ref{subfig:kt5_psiN} which are computed using ground truth rotational operators with added noise with standard deviations of $10^{-3}, 10^{-2}, 10^{-1}$, and $1$. This shows that our method learns operators that are close in structure to the ground truth operators and the performance they achieve is similar to ground truth operators with additive Gaussian noise with a standard deviation of $10^{-2}$. Additionally we see that the depth inference performance with learned operators improves significantly over the first 100-200 training steps but requires fine-tuning for many steps after that to achieve optimal performance.

\subsection{Kinematogram Experiments}\label{sec:chap3_kine}

\begin{figure}[t]

\centering
\begin{subfigure}[b]{0.3\textwidth}
  \centering
	{\includegraphics[width=0.95\textwidth]{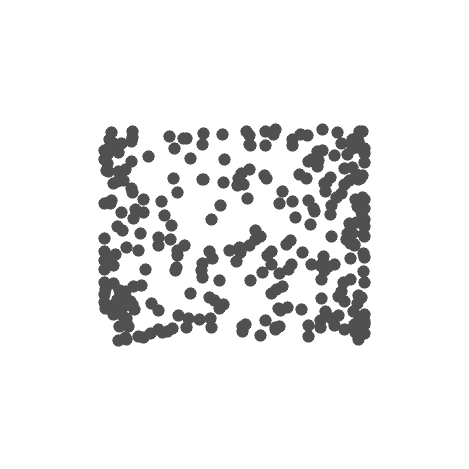}}
  \caption{}
	\label{subfig:kineProj}
\end{subfigure}
\begin{subfigure}[b]{0.4\textwidth}
 \centering
	{\includegraphics[width=0.99\textwidth]{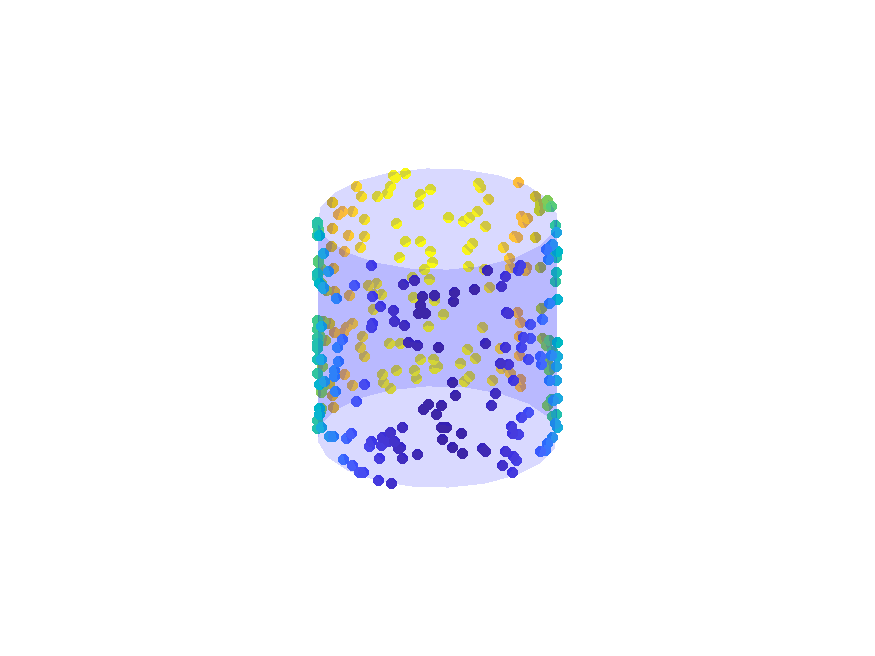}}
 \caption{}
	\label{subfig:kineColor}
\end{subfigure}

  \caption[Visualization of kinematogram visual stimuli.]{\label{fig:kineVis} Visualization of kinematogram visual stimuli. (a) Example of a 2D kinematogram stimulus which is the projection of random dots on the surface of a cylinder. (b) 3D ground truth structure of the points in the kinematogram stimulus. The points are randomly sampled from the cylinder surface and colored by their depth. }
	
\end{figure}

Random dot kinematograms are displays of dots on the surface of or within invisible rotating shapes. Still frames of kinematogram inputs appear as random dots with no perceptible structure (see Fig.~\ref{subfig:kineProj}). However, the motion of the dots elicits the perception of a 3D structure. Fig.~\ref{fig:kineVis} shows the 2D projection of random points along with the 3D structure of the points on the surface of a cylinder. This perception of depth through motion is termed the "kinetic depth effect"~\citep{wallach1953kinetic}. The random dot kinematogram visual stimulus has been used for many structure from motion experiments because it isolates the use of motion cues from the use of other possible depth cues. We use our depth inference model with transport operators learned from 2D projections of rotational motion in order to estimate depths for random points that are located within the volume of invisible rotating shapes. We compare characteristics of our experimental results to the performance of humans on structure from motion tasks with random dot kinematograms.

For these experiments, we create kinematogram stimuli by randomly selecting $N_P$ 3D points within the volume of a cylinder. Sequences are generated by rotating points around the $x$-axis at a rotational speed specified by the angle between frames $\theta$. The points in each frame are ortographically projected to the $xy$-plane. Point correspondences are estimated by pairing nearest neighbors in the projected inputs from one frame to the next using the Hungarian algorithm~\citep{kuhn1955hungarian}. We use the inference procedure described in Section~\ref{sec:chap3_TO_2D} to infer the depth and coefficients.  During inference, we use an inference window of $N_T - 1$ preceding frames to infer depths for the points in the current frame.  We can vary the parameters of the stimuli and the inference procedure and analyze their impact.

\begin{figure}[t]

\centering
\begin{subfigure}[b]{0.28\textwidth}
  \centering
	{\includegraphics[width=0.98\textwidth]{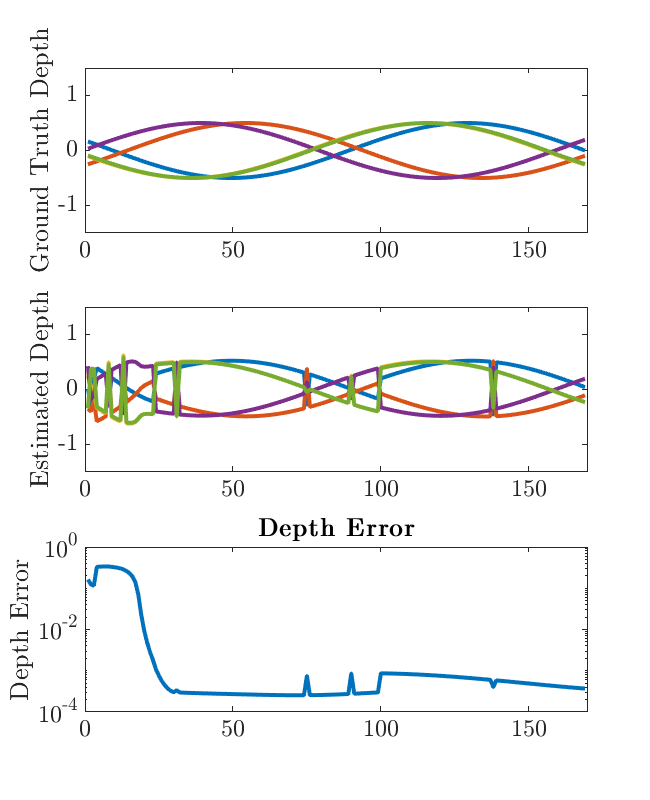}}
  \caption{}
	\label{subfig:kineEstB0}
\end{subfigure}
\begin{subfigure}[b]{0.40\textwidth}
 \centering
	{\includegraphics[width=0.98\textwidth]{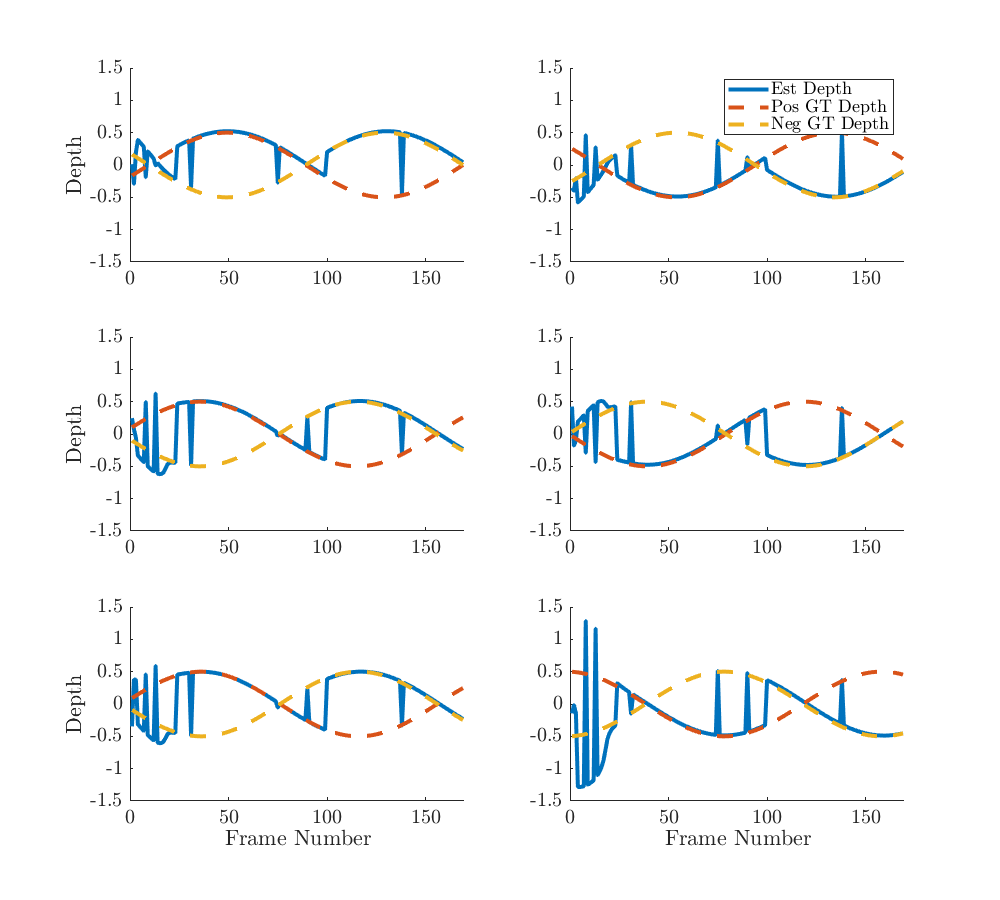}}
 \caption{}
	\label{subfig:kineEstPosNeg}
\end{subfigure}

  \caption[Example of depths inferred for random dots in a kinematogram sequence.]{\label{fig:kineDepthEstB0} Example of depths inferred for random dots in a kinematogram sequence. (a) In the top plot, each line represents the ground truth depth of a single random point over the rotational sequence of the kinematogram. In the middle plot each line represents the estimated depth for the same points as in the top plot. The bottom plot shows the depth error between the estimated and ground truth depths over the sequence. (b) Each plot shows the estimated depth for a single point with the sequences of positive and negative ground truth depths overlaid. }
	
\end{figure}

Fig.~\ref{fig:kineDepthEstB0} shows depths that are inferred for a random dot kinematogram sequence on a cylindrical structure by minimizing the objective in \eqref{eq:DI_obj}. In this experiment, $N_P = 20$, $\theta = 2 \degree$ and $N_T = 30$. Each line in the top and middle plot of Fig.~\ref{subfig:kineEstB0} is the depth for one of five stimulus points. In the early stages of the kinematogram sequence, the number of frames in the inference window is only as large as the number of frames that have appeared (which is less than $N_T$). Once more than $N_T$ frames have appeared, the depth and coefficient inference will make use of only the current frame and the $N_T-1$ preceding frames. This build up in the angular extent of rotation explains the larger depth errors early in the sequence, and we will analyze this further below.

The estimated depth in Fig.~\ref{fig:kineDepthEstB0} has discontinuities that result in the sign of the depth switching. This natural phenomenon is due to the fact that the orthographically projected random dot kinematogram stimulus is a bistable perceptual representation~\citep{andersen1998perception}. That is, it is an ambiguous representation in which there are two correct perceptual structures. All the points could be rotating in a clockwise direction points with a specific combination of positive and negative depths or they could be rotating in a counterclockwise direction with the opposite combination of positive and negative depths. Each of these perceptual estimates is equally correct for the stimulus. Therefore, when computing the error metrics, we correct for the direction of the inferred rotation so it corresponds to the ground truth direction (which is clockwise in all of our experiments). This is done by generating a path with the inferred transformation coefficients and identifying the rotation direction of the points on that path. If the inferred rotation is moving in a counterclockwise direction, we reverse the signs of the depths prior to computing the error metrics. The bottom plot Fig.~\ref{subfig:kineEstB0} shows the depth error for the kinematogram sequence. The depth error is high at the beginning of the sequence due to the limited angular extent of rotation. As the angular extent of rotation increases, the depth error decreases and remains low even while the signs of the depths switch. In Fig.~\ref{subfig:kineEstPosNeg}, we overlay the estimated depth for individual points on top of sequences with both of positive and negative ground truth depth values. This shows, whichever direction of rotation is inferred, the depths are aligned with either the positive or negative ground truth depth values. 

\begin{figure}[t]

\centering

\begin{subfigure}[b]{0.35\textwidth}
  \centering
	{\includegraphics[width=0.95\textwidth]{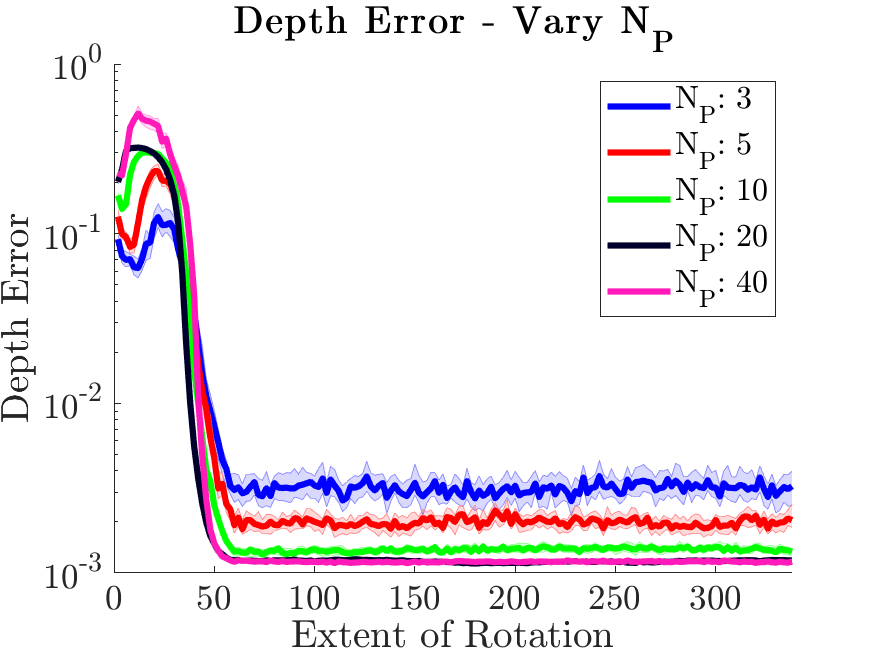}}
  \caption{}
	\label{subfig:kineDepth_ptNum}
\end{subfigure}
\begin{subfigure}[b]{0.35\textwidth}
 \centering
	{\includegraphics[width=0.95\textwidth]{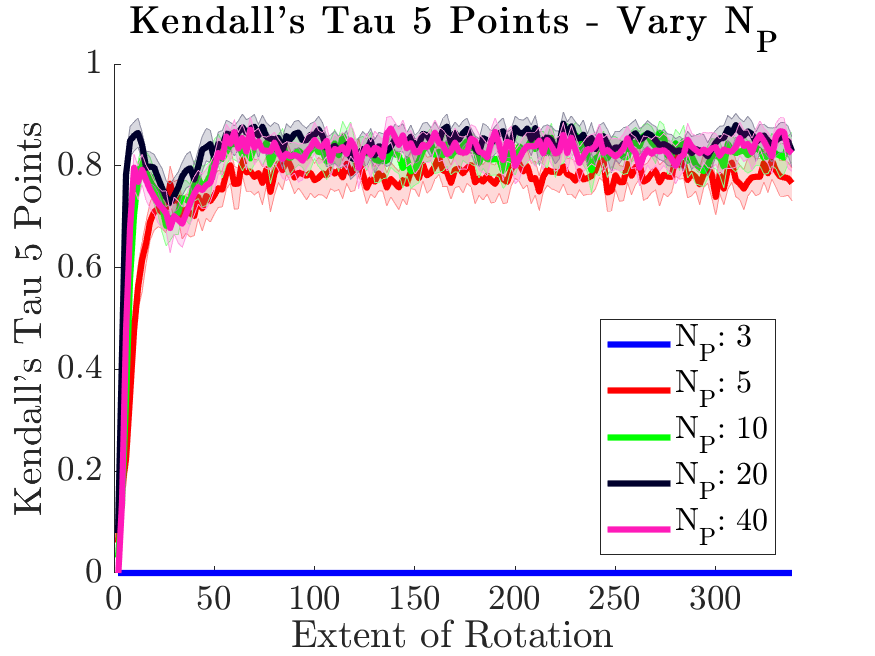}}
 \caption{}
	\label{subfig:kineKT5_ptNum}
\end{subfigure}

\begin{subfigure}[b]{0.35\textwidth}
  \centering
	{\includegraphics[width=0.95\textwidth]{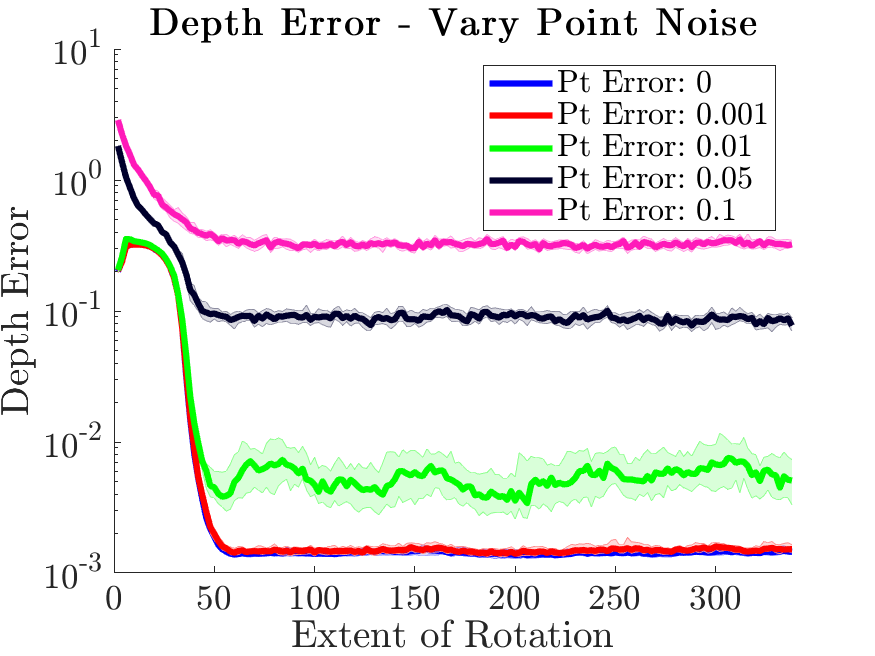}}
  \caption{}
	\label{subfig:kineDepth_ptNoise}
\end{subfigure}
\begin{subfigure}[b]{0.35\textwidth}
 \centering
	{\includegraphics[width=0.95\textwidth]{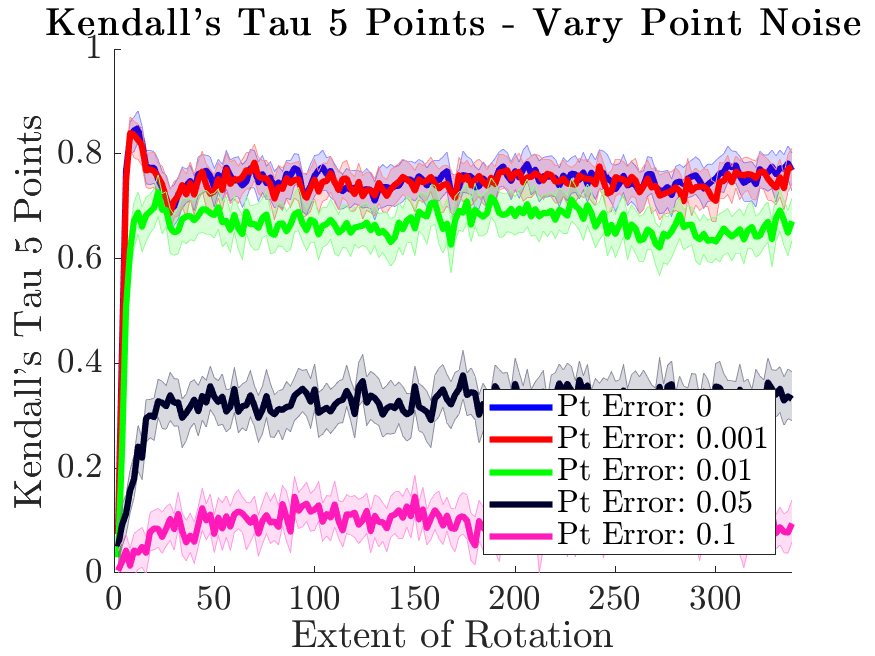}}
 \caption{}
	\label{subfig:kineKT5_ptNoise}
\end{subfigure}

  \caption[Quantitative metrics for random dot kinematogram depth estimates.]{\label{fig:kineQuant} Quantitative metrics for random dot kinematogram depth estimates. Depth error and Kendall's Tau: (a-b) as $N_P$ increases. (c-d) as the standard deviation of noise added to the point locations increases. }
	
\end{figure}

This bistable phenomenon is observed in pyschophysical experiments as well. Specifically, subjects incorrectly identify the rotation direction of orthographically projected stimuli $50.3\%$ of the time~\citep{petersik1979three}. In the experiments shown in Fig.~\ref{subfig:kineDepth_ptNum}, the clockwise rotation is estimated $50.04\%$ of the time.

Fig.~\ref{fig:kineQuant} contains plots demonstrating the performance of our model on random dot kinematogram stimuli as we vary parameters of both the inference algorithm and the kinematogram inputs. These plots compute the depth MSE and the Kendall's Tau for 5 randomly selected dots in the stimulus. Fig.~\ref{subfig:kineDepth_ptNum} and Fig.~\ref{subfig:kineKT5_ptNum} examine the influence of the number of stimulus points\footnote{Note in this experiment we use ground truth correspondences between sample points in each kinematogram frame in order to focus on the impact of $N_P$ on the inferred depth sequence independent of our point correspondence technique.}.   Increasing the number of points improves the accuracy of the depth estimates but does not significantly impact the accuracy of the depth ordering. The Kendall's Tau values for $N_P \geq 10$ have a spike at the beginning of the kinematogram sequence. This is due to the trade-off between the data fidelity term and the depth regularizer term in the objective. With fewer rotational frames in the inference window early in the kinematogram sequence, the optimization results in large magnitudes for the inferred depths that lead to small errors in the data fidelity term but large values for the depth regularizer. Therefore, the ordering of the depths is accurate but the exact depth values are inaccurate because they are off by a scale in magnitude. As the kinematogram sequence continues, with more frames in the inference window, the magnitudes of the depths decrease and reduce the depth regularizer term but this leads to a an increase in the data fidelity term associated less accurate depth ordering. We see this as the depth error decreases (because depth magnitudes are reducing) in conjunction with a decrease Kendall's Tau values. Decreasing the number of points eliminates this spike\footnote{The Kendall's Tau we report in Fig.~\ref{subfig:kineKT5_ptNum} is for five randomly selected stimulus points so the value for $N_P = 3$ is set to zero because there are fewer than five points to use for this metric computation.}. We examine the impact of adding Gaussian noise to the point locations in Fig.~\ref{subfig:kineDepth_ptNoise} and Fig.~\ref{subfig:kineKT5_ptNoise}. 
The depth is consistently accurate with point location noise up to a standard deviation of $10^{-2}$ and depth error increases after that. The introduction of point noise also eliminates the spike in Kendall's Tau at the beginning of the sequence.

This perceptual build up of an accurate estimate of point depths is observed in structure from motion experiments~\citep{hildreth1990perceptual}. \citeauthor{hildreth1990perceptual} performed experiments where they displayed orthographic projections of three points rotating about a central axis and asked subjects to order the depths of the three points. They computed the percentage of correct ordering responses - a metric similar in nature to our Kendall's Tau metric. They found that the percent of correct depth ordering increased as the angular extent of rotation increased up to about $40 \degree$ of rotation after which it plateaued. We observe the same build up and plateau of point ordering accuracy (as judged by Kendalls' Tau). They also observed degradation in performance as Gaussian noise was added to the point locations as we see in Fig.~\ref{subfig:kineKT5_ptNoise}. 

\section{Discussion}

The main contribution of this work is a generative model framework
for learning and inference of 3D manifold-based transformations from 2D projections.   A key innovation of the model is an inference procedure that jointly estimates scene geometry (point depth) and transformation parameters from a sequence of 2D views via gradient descent through a transport operator. Using this procedure, we show that it is possible to learn, without any prior knowledge of transformations or point depth, the dictionary elements that generate rotational motion from 2D projections of rotating points. This model lays the groundwork for explaining the development and adaptation of internal representations of natural variations that are observed in the world. Additionally, our depth inference model enables the investigation of data characteristics that may influence the capacity for accurate depth estimation. This allows us to connect model performance with various data characteristics and algorithmic parameters to human performance on perceptual studies.

An important factor in accurate depth estimation and ordering is that a large angular extent is spanned by the set of input frames used for inference (see Fig.~\ref{fig:quant_rotExt} and Fig.~\ref{fig:kineQuant}). This supports the notion that humans build up their perception of 3D structure during random dot kinematogram rotation sequences~\citep{hildreth1990perceptual}. We also show that increases in the number of random dot stimuli result in improvement in depth inference performance (see Fig.~\ref{subfig:depthErr_numPts}, Fig.~\ref{subfig:kt5_numPts}, Fig.~\ref{subfig:kineDepth_ptNum}, and Fig.~\ref{subfig:kineKT5_ptNum}). This connects to kinematogram experiments that indicate a greater number of coherently transforming points results in a stronger depth percept from moving points~\citep{todd1988apparent,dosher1989ratings,sperling1989kinetic,braunstein1987minimum}. Our model also demonstrates the same direction switching phenomenon with the bistable kinematogram stimulus that humans perceive~\citep{petersik1979three}. 

\subsection{Psychophysical implications}
Our model has the flexibility to adapt to many different test scenarios that are inspired by human performance on mental rotation and structure from motion tasks. In our experimentation, we tuned parameters like the inference window length $N_T$, the angle between frames $\theta$, and the number of stimulus points $N_P$ to achieve the most accurate depth estimates. However, experiments show, even when humans perceive the correct shape, they often have inaccurate estimates of depth magnitude ~\citep{todd1988apparent}, especially when viewing limited numbers of transforming points~\citep{dosher1989ratings}. From our model performance, this may suggest that humans rely on a smaller angular extent of rotation for inferring depth or that they do not utilize a prior on expected depths. In the future, we can vary our model parameters to explore comparisons with potentially inaccurate human depth estimation in various tests settings.

In this work we do not directly relate the internal rotation model developed here to the rich area of mental rotation experiments. In particular, the seminal work in that area suggests a monotonically increasing relationship between rotation angle between views of an object and the human reaction time~\citep{shepard1971mental}. The manifold based model presented here may have a similar connection between processing time and rotation angle because the transport operators can generate transformations similar to the internal representation of 3D rotations described by humans in these studies. It may be a fruitful to examine the performance of our transformation model on mental rotation tasks and to compare to human performance on similar tasks.

\subsection{Future Improvements}

Ultimately, addressing the underlying neural mechanisms of 3D perception 
will require formulating a more biologically plausible model. The inference and learning in the current model are performed using quasi-Newton optimization and gradient descent, respectively. The optimization objective is non-convex and does not naturally lend itself to a parallel representation similar to neural architectures. Moving forward, we suggest developing an optimization procedure that is more biologically plausible. 

As our focus in this work is on the development of a transformation learning framework, we assume ground truth point correspondences for the learning and inference experiments in Section~\ref{sec:chap3_DIExp} through Section~\ref{sec:chap3_TO_learn}. However, identifying point correspondences from different views of the same scene is a challenging task and one that has been a focus of many computer vision algorithms~\citep{ullman1979interpretation,fischler1981random,zbontar2016stereo,luo2016efficient}. Going forward, incorporating point correspondence estimates into this framework will lead to a more versatile, biologically plausible model.

Finally, a step towards biological plausibility is extending this model to be robust to incoherent point motion and additional moving objects. An initial approach to improving the robustness to incoherent motion is to employ random sample consensus (RANSAC)~\citep{fischler1981random}. With this method, transport operator coefficients could be estimated from random subsets of points in the scenes and the final transformation parameters could be chosen as those that explain the transformation between the largest number of random subsets of points.

\bibliographystyle{apalike}

\bibliography{references.bib}

\pagebreak
\appendix

\section{Egocentric vs Allocentric}\label{appsec:egoAllo}

\begin{figure}[t]

\centering
\begin{subfigure}[b]{0.32\textwidth}
  \centering
	{\includegraphics[width=0.7\textwidth]{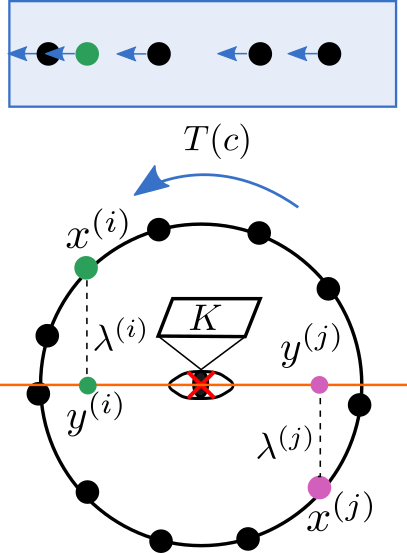}}
  \caption{}
	\label{fig:alloSetup}
\end{subfigure}
\begin{subfigure}[b]{0.32\textwidth}
 \centering
	{\includegraphics[width=0.95\textwidth]{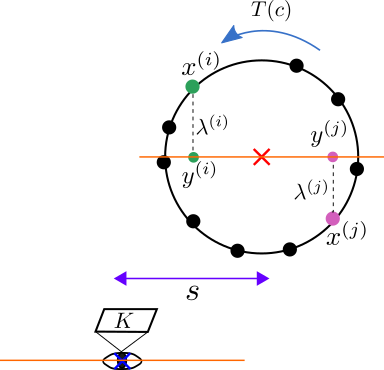}}
 \caption{}
	\label{subfig:originOff}
\end{subfigure}

  \caption{Top-down views of the depth inference problem setup in an allocentric framework where the object rotates around the observer. The 3D points $\vtr{x}^{(i)}$ have an associated depth $\lambda^{(i)}$.  Each point is projected onto the orange viewing plane using the projection matrix $\mtx{K}$. This results in the 2D projected points $\vtr{y}^{(i)}$. (a) Example when the origin of the motion is located at the viewer location. The 3D points are rotating counter-clockwise around the axis and the points in the blue shaded box on top indicate the direction of motion of the projected points. (b) Example when the origin of motion is offset from the viewer location.  }
	
\end{figure}

The model presented can be applied to two viewing frameworks -- the allocentric framework in which points rotate around the observer and the egocentric framework in which the observer rotates with respect to the surrounding world. When the motion is centered around the origin (i.e., when the origin of the observer coordinate system and the world coordinate system is the same), the allocentric and egocentric learning frameworks utilize the exact same model. Comparing Fig.~\ref{subfig:egoSetup} and Fig.~\ref{fig:alloSetup}, the only difference between the allocentric and egocentric frameworks when motion is centered at the origin is the direction in which the projected points move with respect to the rotational motion. When the viewer rotates in a counter-clockwise direction in the egocentric framework, the projected points with positive depth (i.e. points in front of the viewer) move to the right in the viewing plane. On the other hand, when the points rotate in a counter-clockwise direction in the allocentric framework, the projected points with positive depth move to the left in the viewing plane.  Therefore, the model and experiments can correspond interchangeably to the egocentric or allocentric frameworks. 

For the work presented here, we are assuming that the motion generated by the transformations we wish to learn is centered at the viewer location. However, if the origin of transformational motion is offset from the viewer location (example shown in Fig.~\ref{subfig:originOff}), this model can be easily extended to incorporate an origin offset. This offset may be known a priori or estimated by viewing the point motion over several frames. 

\section{Depth-Angle Ambiguity}\label{appsec:depthAng}

\begin{figure}[ht]

  \centering
	{\includegraphics[width=0.15\textwidth]{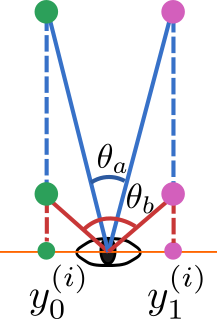}}

  \caption[Example of depth-angle ambiguity.]{\label{fig:depthAng} Example of depth-angle ambiguity. Points $\vtr{y}_0^{(i)}$ and $\vtr{y}_1^{(i)}$ are projections of a 3D point that is transformed from view 0 (green points) to view 1 (pink points). These points could result from a small rotation angle $\theta_a$ on points with large depth magnitudes or a larger rotation angle $\theta_b$ with smaller depth magnitudes.}
	
\end{figure}

Fig.~\ref{fig:depthAng} shows a visualization of the depth-angle ambiguity that exists with orthographic projection. When viewing projected points $\vtr{y}_0^{(i)}$ and $\vtr{y}_1^{(i)}$ from two separate views, these points could be either projections of points with large depths that undergo rotation with a smaller angle $\theta_a$ or points with small depths that undergo rotation with a larger angle $\theta_b$. 

\section{Visualization of Operator Noise}

\begin{figure}[ht]

  \centering
	{\includegraphics[width=0.97\textwidth]{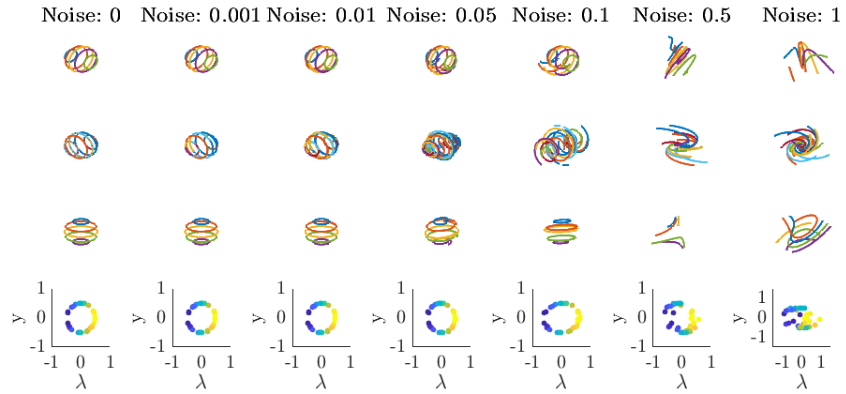}}

  \caption[Examples of noisy operator trajectories.]{\label{fig:psiNoiseEx} Examples of noisy operator trajectories. Each column shows examples of the ground truth rotational operators with additive Gaussian noise with increasing standard deviation. Rows 1-3 show the trajectories for the three rotational operators. Row 4 shows depth inferred for points projected from a rotating cylinder using the three operators in each column. The operators do not vary much in appearance from the ground truth operators (first column) with noise standard deviations of 0.01 or less. For noise standard deviations larger than that, the operators diverge from rotational operators and the point depths no longer look like a cylinder.}
	
\end{figure}

Fig.~\ref{subfig:depthErr_psiN} and Fig.~\ref{subfig:kt5_psiN} show the quantitative impact that adding noise to ground truth rotation operators has on the accuracy of inferred depth. To provide an intuitive understanding of the effect of noise on the operators, \ref{fig:psiNoiseEx} shows the example trajectories for operators with increased noise standard deviation.

\section{Model Analysis}

\begin{figure}[t]

\centering
\begin{subfigure}[b]{0.32\textwidth}
  \centering
	{\includegraphics[width=0.95\textwidth]{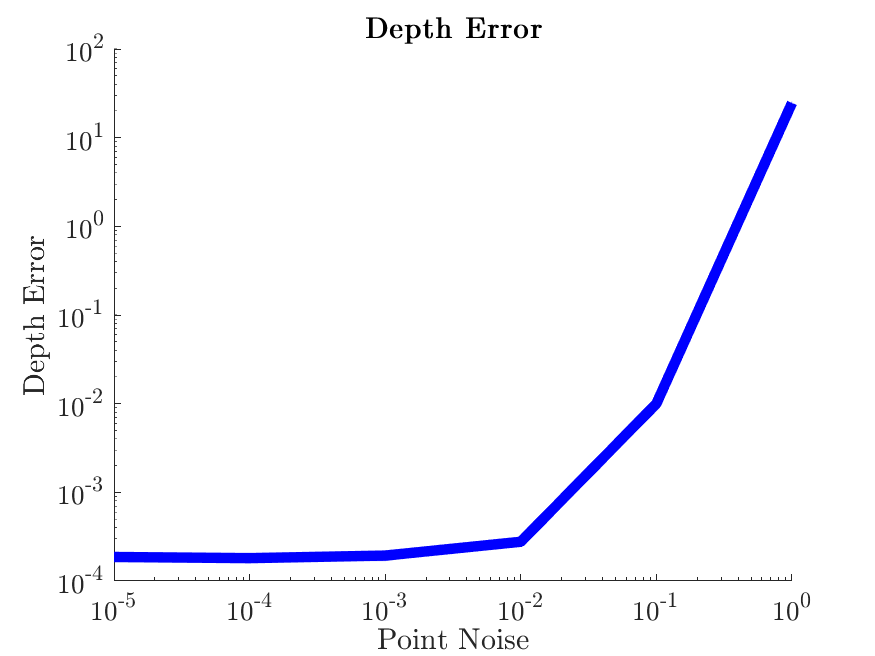}}
  \caption{}
	\label{subfig:depthErr_ptN}
\end{subfigure}
\begin{subfigure}[b]{0.32\textwidth}
 \centering
	{\includegraphics[width=0.95\textwidth]{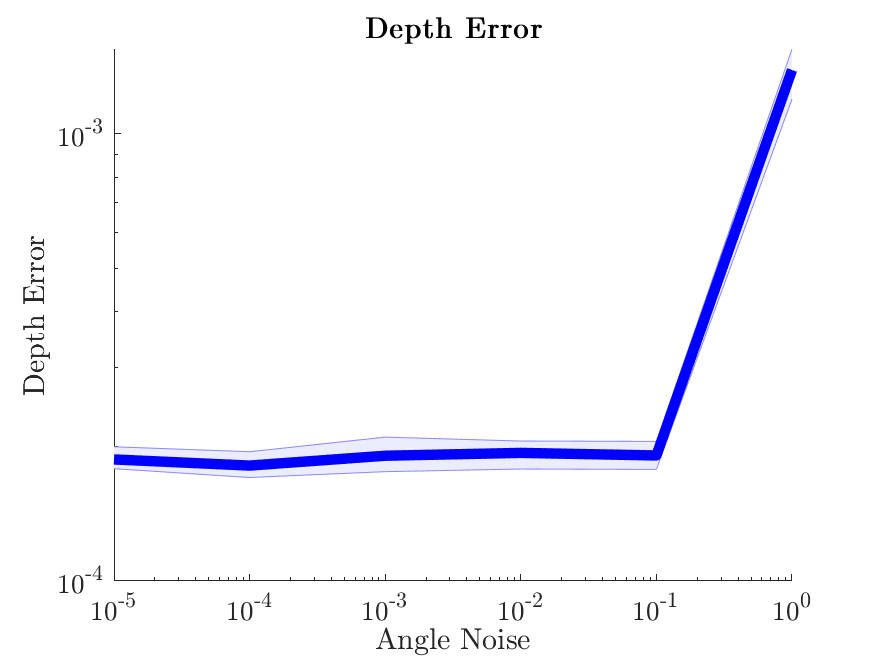}}
 \caption{}
	\label{subfig:depthErr_angN}
\end{subfigure}
\begin{subfigure}[b]{0.32 \textwidth}
 \centering
	{\includegraphics[width=0.95\textwidth]{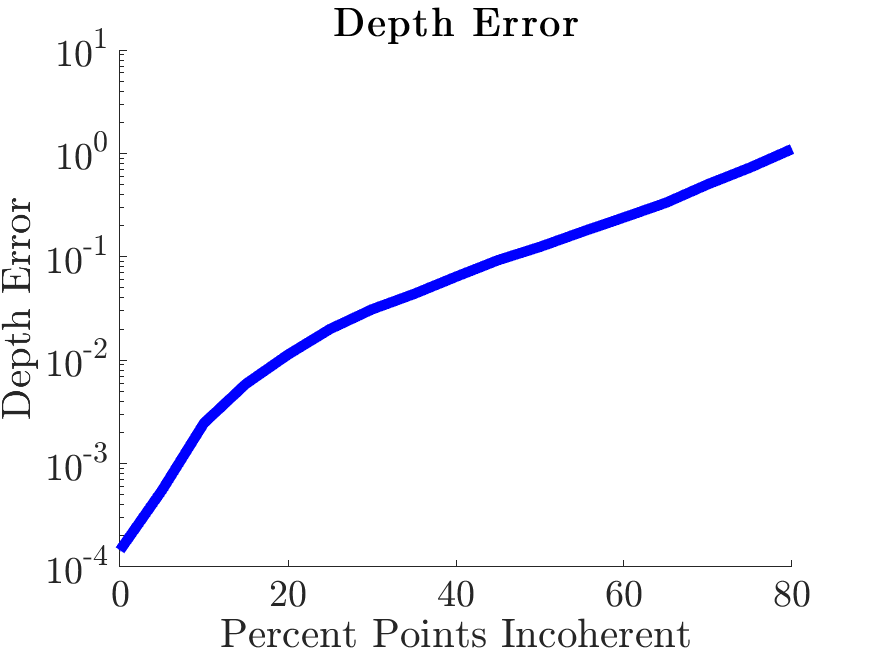}}
 \caption{}
	\label{subfig:depthErr_inco}
\end{subfigure}

  \caption[Depth error in the presence of mismatch between the model assumptions and stimulus characteristics.]{\label{fig:quant_noise} Depth error for depth inference in the presence of mismatch between the model assumptions and stimulus characteristics. (a) Depth error as the standard deviation of additive Gaussian noise in the 2D point locations increases. The model is robust to noise with a standard deviation up to around $10^{-2}$. (b) Depth error as we increase the standard deviation of the Gaussian noise added to the magnitude of the rotation angles between frames in the inference window $\theta$. The model is robust to noise with a standard deviation up to around $10^{-1}$. (b) Depth error as the percent of incoherently moving points is increased. The depth error quickly increases after $5 - 10\%$ of points are incoherent. }
	
\end{figure}

In this section we will detail our analysis of the robustness of our model to a mismatch between the model assumptions and data characteristics.  The first assumption is that we have ground truth keypoints locations. We analyze the model robustness to noise in the keypoints locations when using ground truth rotational operators by adding Gaussian noise with progressively larger standard deviations to the 2D point locations and observing the variations in the depth error (Fig.~\ref{subfig:depthErr_ptN}). For reference, points in these experiments have values in the range of $[-1,1]$. Results indicate that the model is robust to noise with a standard deviations of up to around $10^{-2}$.

Next we analyze the assumption that each frame in the inference window results from rotations with the same speed and direction as preceding frames. In natural systems, objects are unlikely to undergo rotation at the exact same speed over many frames. To test the effect of variations in the rotation speed, we create rotation sequences in which each frame uses the same axis of rotation but the magnitude of the rotation angle is varied by adding Gaussian noise with a standard deviation scaled by the angle of rotation\footnote{The angles used in this experiment have a mean of $2\degree$.}.  In this setting, we infer the depth and coefficients using ground truth rotational operators and compute the depth error shown in Fig.~\ref{subfig:depthErr_angN}. Our model is robust to variations in rotation speed up to a standard deviation of around $10^{-1}$. 

Our final assumption in the experiments in this paper is that all the points are moving coherently. We analyze the performance of our model in the presence of points that are moving incoherently with the rotating points. Fig.~\ref{subfig:depthErr_inco} shows the depth error as we increase the percentage of incoherently moving points when $N_P = 100$. This shows that incoherent points significantly impact the accuracy of the 3D structure estimate.

\section{Inference Details}\label{appsec:DIInferDet}
We perform depth inference in Section~\ref{sec:chap3_DIExp} and \ref{sec:chap3_TO_learn} using the objective in \eqref{eq:DI_obj}. In the inference experiments in Section \ref{sec:chap3_DIExp}, we set $\gamma = 0.1$, $\zeta = 0.01$, and $\beta = 10^{-3}$. These parameters are selected because they yield accurate inferred depth estimates. We add Gaussian noise with a standard deviation of $10^{-3}$ to the ground truth operators in most experiments in Section \ref{sec:chap3_DIExp}. Unless otherwise stated, we set $N_P = 20$, $\theta = 2 \degree$, and $N_T = 30$. For the quantitative analysis of inference on many trials, we use inputs points that are 2D projections of random 3D points that are undergoing rotation about a randomly selected 3D axis. The rotation angle used for a given trial is sampled from a distribution with a mean of $\theta$ and a standard deviation of $0.516 \degree$.

Because the training objective is nonconvex, optimization may result in local minima. To avoid resulting in local minima, we perform inference for the same inference window several times using several random restarts. That is, we randomly sample a new intialization and infer coefficients and depths using that starting point. This often results in different final inferred outputs. We choose the inferred output associated with the lowest final objective from inference. For the experiments in Section \ref{sec:chap3_DIExp}, we use five random restarts.

For the inference experiments, we compute the mean squared error between the ground truth depths and estimated depths and the Kendall's Tau between the ordering of the truth depths and estimated depths. As mentioned in Section \ref{sec:chap3_kine}, the kinematogram stimulus is bistable which means it can result in two separate percepts (i.e., clockwise rotation or counter-clockwise rotation). We observe switching in inferred direction and signs of the depths with our model and account for that when computing the depth inference metrics. Specifically we generate a path with the transformation defined by the inferred coefficients and observe the direction of motion of that path to determine which direction of rotation we inferred. If the inferred rotation is counter-clockwise then we multiply the depths by $-1$ because the rotation of the points in ground truth sequence is clockwise.

\section{Training Details}\label{appsec:DITrainDet}

We train the transport operators using gradient descent. The operators are intialized with Gaussian noise with a standard deviation of 0.3. The training run that resulted in the operators used in the kinematogram experiments used the parameters in Table~\ref{tab:DI_Trainparams}.

\begin{table}[!htb]
\centering
\caption[Training parameters for learning rotational operators from 2D projected inputs.]{Training parameters for learning rotational operators from 2D projected inputs.}
\label{tab:DI_Trainparams}
\begin{tabular}{||l||} 
 \hline
 Training Parameters \\ 
 \hline
 $M:$ 3 \\
 $lr_{\text{begin}} : 0.5$  \\
 $\zeta:$ 0.1  \\
 $\gamma:$ 0.15   \\
$\beta$: $10^{-4}$ \\
$\theta: 10\degree$ \\
$N_T: 20$ \\
$N_P: 20$ \\
Training Steps: 10,000 \\
   number of restarts for coefficient inference: 25 \\
 \hline
\end{tabular}

\end{table}

We begin training at a specific learning rate and increase it if there is a successful learning step (i.e., one that decreases the learning objective) or decrease it if there is a failed learning step. While large learning rates aid in efficient gradient steps in the beginning of training, we find that decreasing the learning rate consistently towards the end of training leads to more stable final transport operator representations. Specifically, we start decreasing the learning rate at 3000 training steps and decay it by a multiplication factor of 0.9997 at each step.

\begin{figure}[ht]

  \centering
	{\includegraphics[width=0.97\textwidth]{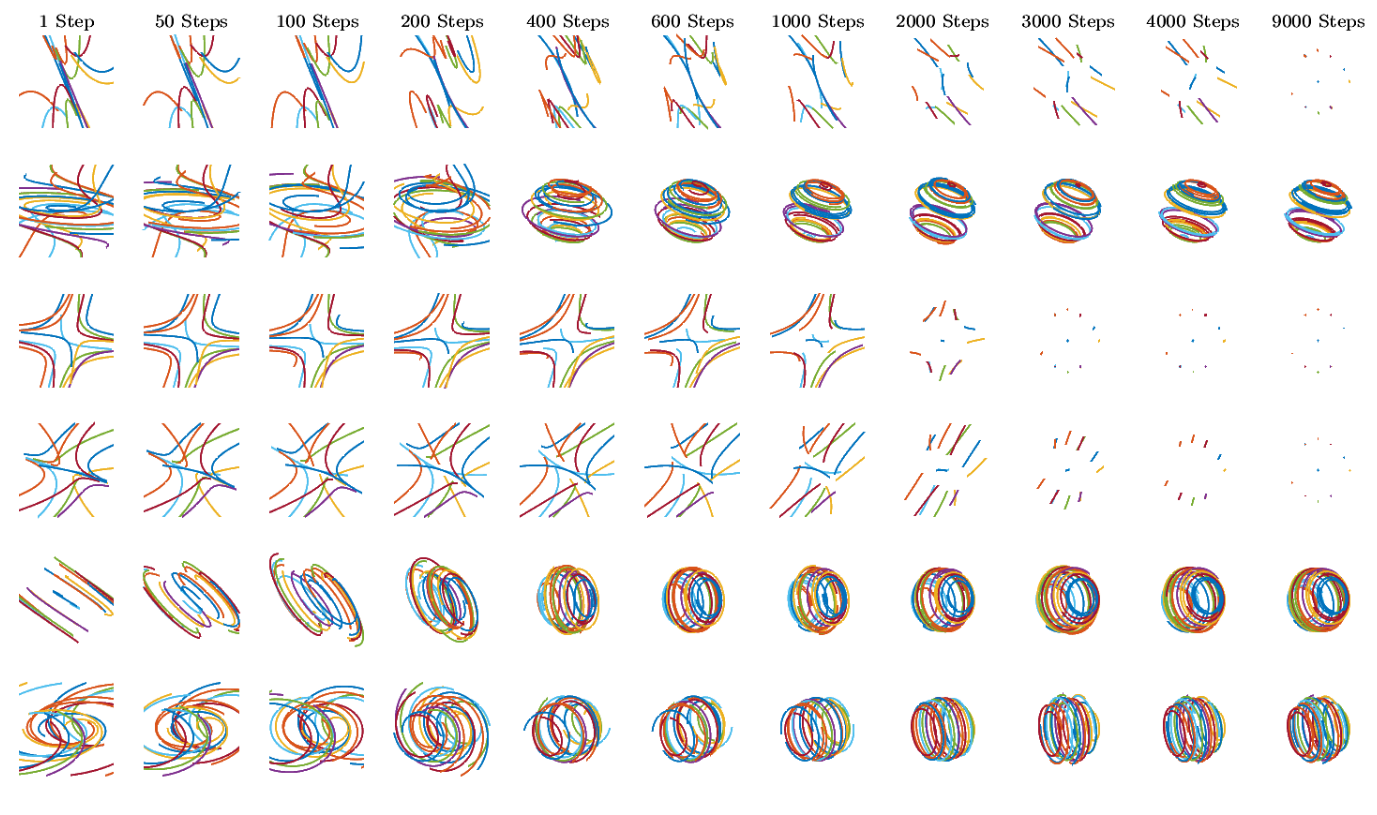}}

  \caption[Transport operator trajectories during training with six dictionary elements.]{\label{fig:trainTraj6} Transport operator trajectories during training. Each row represents one of the six learned operators. Each column shows the trajectories at a different training step. The operators begin with random initializations at step 1 and reach a rotation structure around 400 steps. After that, the three operators that do not represent rotation have their magnitudes reduced because they are not being used. At the end of training, there are three operators with clear rotational structure.}
	
\end{figure}

Fig.~\ref{fig:trainTraj6} shows the trajectories of operators learned when $M = 6$. This highlights the usefulness of the Frobenius norm regularizer on the dictionary elements. If a transport operator is not being used for representing manifold paths, then its magnitude is reduced to nearly zero. Training with six operators utilizes the following parameters: $\zeta = 0.1,$ $\gamma = 0.08,$ $\beta = 10^{-4},$ $\theta = 10 \degree,$ $N_T = 20,$ $N_P = 20,$ 10000 training steps, and 25 random restarts.

\section{Kinematogram Experimental Details}\label{appsec:DIKineDet}

For the kinematogram experiments, we use a base set of parameters which are shown in Table \ref{tab:DI_Kineparams}. For individual experiments, we vary subsets of the parameters from these baseline values. As with the other experiments, we correct for depth sign switching before computing the error metrics for the kinematogram tasks.

\begin{table}[!htb]
\centering
\caption[Parameters for random dot kinematogram experiments.]{Parameters for random dot kinematogram experiments.}
\label{tab:DI_Kineparams}
\begin{tabular}{||l||} 
 \hline
 Kinematogram Inference  Parameters \\ 
 \hline
 $\zeta:$ 0.01  \\
 $\gamma:$ 0.1   \\
$\beta$: $0$ \\
$\theta: 2\degree$ \\
$\xi: 0$ \\
$N_T: 30$ \\
$N_P: 20$ \\
   number of restarts for coefficient inference: 5 \\
 \hline
\end{tabular}

\end{table}

\section{Kinematogram Dynamic Regularization}

Our model has the capability of reducing the amount of depth sign switching and transformation direction switching by adding a dynamic regularizer during inference. We assume that there is a constant speed of rotation and encourage the transport operator coefficients to be similar from one frame to the next. Note that we already utilize this coefficient consistency in the inference procedure by inferring the same coefficients for all frames in the inference window. The additional regularizer encourages the set of coefficients to be similar from one inference window to the next. This amounts to adding a term to the inference objective that incorporates the previous coefficient estimate:
\begin{equation}\label{eq:conReg}
\begin{split}
    L = \frac{1}{2N_T}\sum_{n=1}^{N_T}\sum_{i=1}^{N_P}\left[\|\vtr{y}^{(i)}_{N_T-n} - \mtx{K}\mtx{T}(n\vtr{c})\widehat{\vtr{x}}^{(i)}_{N_T}(\vtr{\lambda})\|_2^2\right] + \zeta\|\vtr{c}\|_1 \\ + \frac{\beta}{2}\|\vtr{\lambda}\|^2_2 + \frac{\xi}{2}\|\vtr{c}-\vtr{c}_{\text{prev}}\|^2_2.
    \end{split}
\end{equation}

\begin{figure}[t]

 \centering
	{\includegraphics[width=0.3\textwidth]{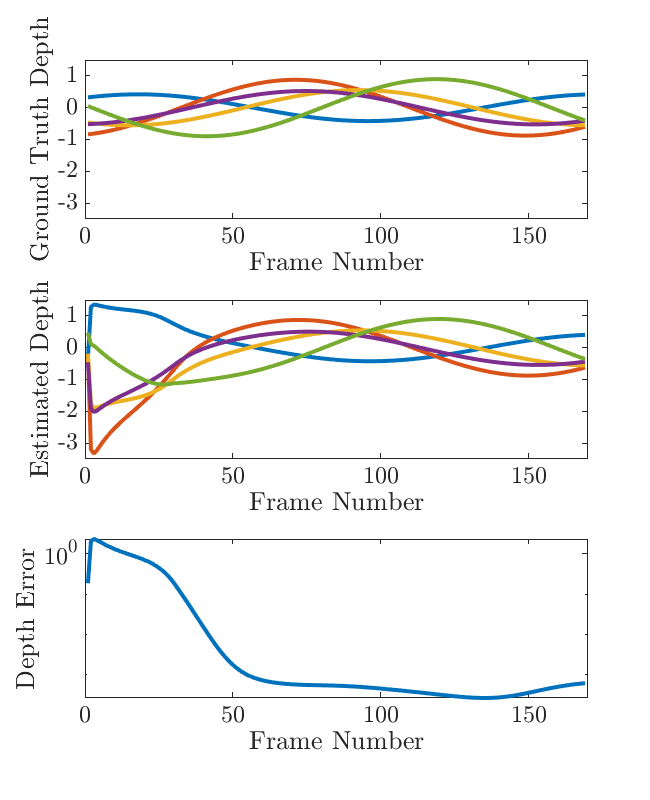}}

  \caption{\label{fig:kineDepthEstB20} Example of depths inferred for random dots in a kinematogram sequence with a dynamic regularizer.  The dynamic regularizer removes the sign switching of the depth values observed in Fig.~\ref{fig:kineDepthEstB0} but it also leads to larger error early in the kinematogram sequence.}
	
\end{figure}

With the addition of the dynamic regularizer on the coefficients, we infer smooth depth estimates for kinematogram sequences. Fig.~\ref{fig:kineDepthEstB20} shows the estimated depths and depth error with dynamic regularization. This regularization eliminates sign-flipping but it also impacts the depth error in the beginning frames of the kinematogram sequence. The initial depth estimates are less accurate because they use only a small window of frames for depth inference and the coefficient regularizer encourages coefficient estimates in later frames to be similar to the initial inaccurate estimates. The depth estimates eventually achieve low error as the inference window gets larger.

\begin{figure}[t]

\centering
\begin{subfigure}[b]{0.35\textwidth}
  \centering
	{\includegraphics[width=0.95\textwidth]{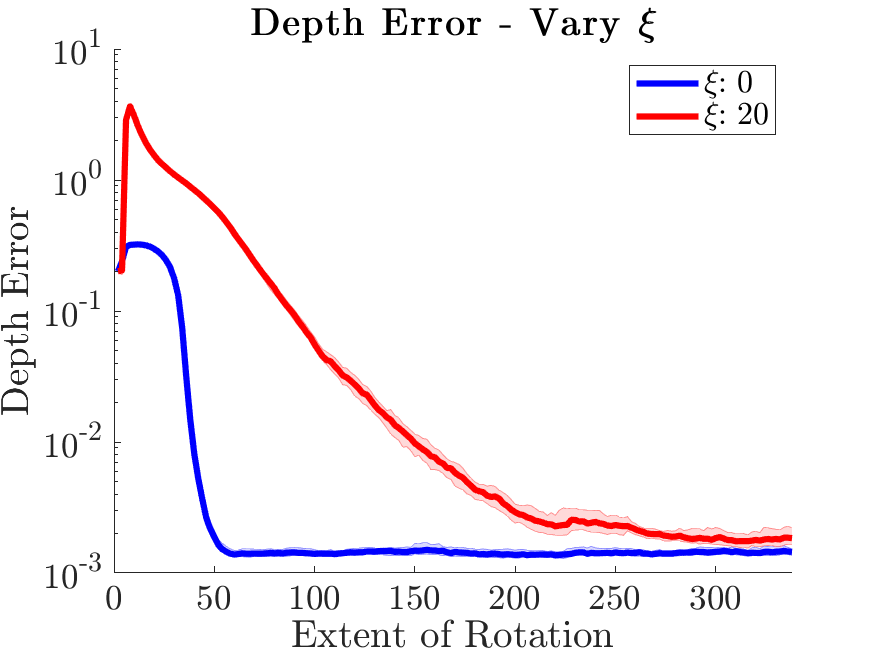}}
  \caption{}
	\label{subfig:kineDepth_xi}
\end{subfigure}
\begin{subfigure}[b]{0.35\textwidth}
 \centering
	{\includegraphics[width=0.95\textwidth]{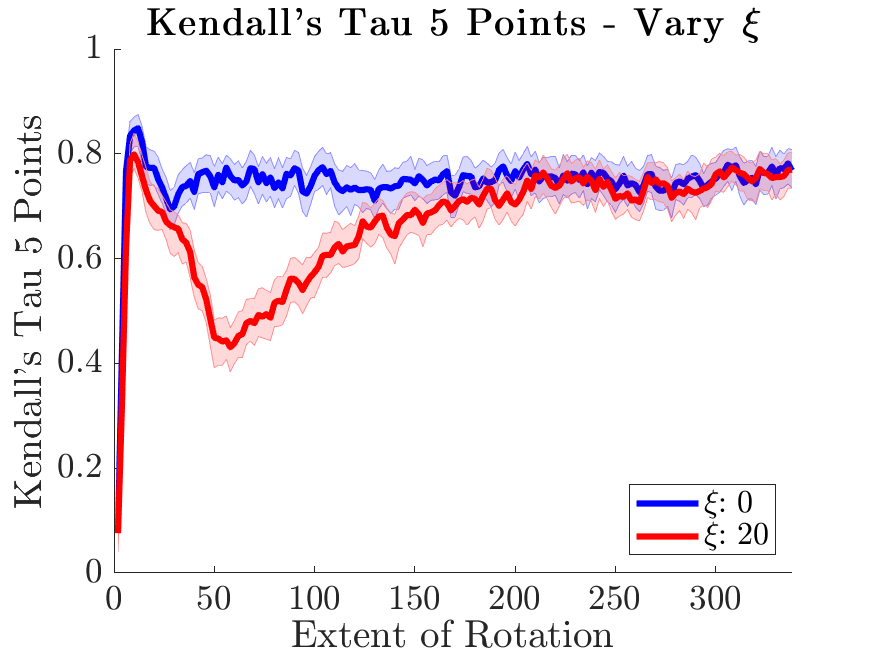}}
 \caption{}
	\label{subfig:kineKT5_xi}
\end{subfigure}

\caption{\label{fig:varyXi}Quantitative metrics for random dot kinematogram depth estimates with an without dynamic regularization.}
\end{figure}

The plots in Fig.~\ref{fig:varyXi} compute the depth MSE for all $N_P$ points and the Kendall's Tau for 5 randomly selected dots in the stimulus. Fig.~\ref{subfig:kineDepth_xi} and Fig.~\ref{subfig:kineKT5_xi} each show the impact of including the dynamic regularizer from \ref{eq:conReg}. When $\xi = 0$, the median depth error looks the same as in Fig.~\ref{fig:kineDepthEstB0}, where it begins high due to the limited angular extent of rotation and decreases as the kinematogram sequence continues.  The Kendall's Tau values with both  $\xi = 0$ and $\xi = 20$ have a spike at the beginning of the kinematogram sequence. Without the dynamic regularizer, the Kendall's Tau value plateaus and remains around the same value until the end of the sequence. As we saw in Fig.~\ref{fig:kineDepthEstB20}, the integration of the dynamic regularizer on the coefficients leads to a delay in achieving optimal accuracy for depth estimates measured by both the depth MSE and Kendall's Tau.

\end{document}